\pdfoutput=1
\documentclass[prx,aps,twocolumn,superscriptaddress,floatfix,preprintnumbers]{revtex4-1}
\usepackage{mathrsfs}
\usepackage{amsfonts}
\usepackage{amsmath,amssymb,bm} 
\usepackage{array}
\usepackage{verbatim}
\usepackage{epsfig}
\usepackage{latexsym}
\usepackage{graphicx,graphics,color} 
\usepackage{hyperref}
\hypersetup{colorlinks, linkcolor = [rgb]{0,0.0,0.5}, citecolor = [rgb]{0,0.0,0.5}, urlcolor = [rgb]{0,0.0,0.5}}
\usepackage{url}
\usepackage[normalem]{ulem}
\usepackage{xcolor}
\usepackage{ulem}
\usepackage{longtable}
\usepackage{multirow}
\usepackage{hhline,multirow,tabularx}
\usepackage{dcolumn}
\usepackage{slashed}
\usepackage{rotating}
\usepackage{mathrsfs}
\usepackage{xr}
\usepackage{soul}
\usepackage{orcidlink}
\usepackage{physics} 
\usepackage{xspace}

\newcommand{\geleven}{{\tt g11}\xspace}

\begin{document}

\preprint{JLAB-THY-23-3881}

\title{{Toward a generative modeling analysis of CLAS exclusive $2\pi$ photoproduction}}

\newcommand{\catania}{INFN Sezione di Catania, I-95123 Catania, Italy}
\newcommand{\genova}{INFN sezione di Genova, I-16146 Genova, Italy}
\newcommand{\unige}{Università di Genova, I-16146 Genova, Italy}
\newcommand{\ceem}{Center for Exploration of Energy and Matter, Indiana  University, Bloomington, Indiana 47403, USA}
\newcommand{\indiana}{Department of Physics, Indiana  University, Bloomington, Indiana 47405, USA}
\newcommand{\jlab}{Thomas Jefferson National Accelerator Facility, Newport  News, Virginia 23606, USA}
\newcommand{\ut}{Institute for Theoretical Physics, T\"ubingen University, D-72076 T\"ubingen, Germany}
\newcommand{\ur}{Institute for Theoretical Physics, Regensburg University, D-93040 Regensburg, Germany}
\newcommand{\msu}{Skobeltsyn Institute of Nuclear Physics, Lomonosov Moscow State University, 119234 Moscow, Russia}
\newcommand{\msuf}{Faculty of Physics, Lomonosov Moscow State University, 119991 Moscow, Russia}
\newcommand{\messina}{Dipartimento di Scienze Matematiche e Informatiche, Scienze Fisiche e Scienze della Terra,
Universit\`a degli Studi di Messina, I-98166 Messina, Italy}
\newcommand{\odu}{Department of Computer Science, Old Dominion University, Norfolk, Virginia 23529, USA}
\newcommand{\bu}{Department of Computer Science, College of Computer Science and Information Technology, Al-Baha University, Al-Baha, Alaqiq 65779, Saudi Arabia}
\newcommand{\davidsonph}{Department of Physics, Davidson College, Davidson, North Carolina 28035, USA}
\newcommand{\davidsoncs}{Department of Mathematics and Computer Science, Davidson College, Davidson, North Carolina 28035, USA}
\newcommand{\adelaide}{CSSM and CDMPP, Department of Physics, University of Adelaide, 5005 Australia}
\newcommand{\agh}{AGH University of Krakow, Faculty of Physics and Applied Computer Science, PL-30-059 Krak\'ow, Poland}

\author{T.~\surname{Alghamdi}\orcidlink {0000-0002-5640-3824}}
\email{talgh001@odu.edu}
\affiliation{\odu} 
\affiliation{\bu} 
\author{Y.~\surname{Alanazi}}
\affiliation{\jlab}
\author{M.~\surname{Battaglieri}\orcidlink{0000-0001-5002-8771}}
\affiliation{\genova}
\author{\L.~\surname{Bibrzycki}\orcidlink{0000-0002-6117-4894}}
\affiliation{\agh}
\author{A.~V.~\surname{Golda}}
\affiliation{\msuf}
\author{A.~N.~\surname{Hiller~Blin}\orcidlink{0000-0002-6854-6259}}
\affiliation{\ut}
\author{E.~L.~\surname{Isupov}}
\affiliation{\msu}
\author{Y.~\surname{Li}}
\affiliation{\odu}
\author{L.~\surname{Marsicano}\orcidlink{0000-0002-8931-7498}}
\affiliation{\genova}
\author{W.~\surname{Melnitchouk}\orcidlink{0000-0002-9521-5973}}
\affiliation{\jlab}
\affiliation{\adelaide}
\author{V.~I.~\surname{Mokeev}\orcidlink{0000-0002-4557-1320}}
\affiliation{\jlab}
\author{G.~\surname{Monta\~na}\orcidlink{0000-0001-8093-6682}}
\affiliation{\jlab}
\author{A.~\surname{Pilloni}\orcidlink{0000-0003-4257-0928}}
\affiliation{\messina}
\affiliation{\catania}
\author{N.~\surname{Sato}\orcidlink{0000-0002-1535-6208}}
\affiliation{\jlab}
\author{A.~P.~\surname{Szczepaniak}\orcidlink{0000-0002-4156-5492}}
\affiliation{\jlab}
\affiliation{\indiana}
\affiliation{\ceem}
\author{T.~\surname{Vittorini}\orcidlink{0009-0002-4390-5670}}
\affiliation{\genova}
\affiliation{\unige}

\begin{abstract}
AI-supported algorithms, particularly generative models, have been successfully used in a variety of different contexts. 
In this work, we demonstrate for the first time that generative adversarial networks (GANs) can be used in high-energy experimental physics to unfold detector effects from multi-particle final states, while preserving correlations between kinematic variables in multidimensional phase space. 
We perform a full closure test on two-pion photoproduction pseudodata generated with a realistic model in the kinematics of the Jefferson Lab CLAS \geleven experiment. 
The overlap of different reaction mechanisms leading to the same final state associated with the CLAS detector's nontrivial effects represents an ideal test case for AI-supported analysis.
Uncertainty quantification performed via bootstrap provides an estimate of the systematic uncertainty associated with the procedure. 
The test demonstrates that GANs can reproduce highly correlated multidifferential cross sections even in the presence of detector-induced distortions in the training datasets, and provides a solid basis for applying the framework to real experimental data.
\end{abstract}

\date{\today}
\maketitle

\section{Introduction}

Photoproduction of two pions, with photon energies in the few-GeV range, is an important process in hadron spectroscopy.
It has been widely used to address 
several fundamental quests, 
such as the `missing baryons' problem, and to demonstrate that
multiparticle final states are necessary to determine the spectrum. 
While copious data are available for single-pion photoproduction, and the correspondent phenomenology is well understood, the addition of a third particle in the final state makes the description of this reaction considerably more complicated. 
At fixed photon energy, the unpolarized single-pion photoproduction cross section is described by a single independent variable, while for two pions three additional variables are needed.
At beam energies of a few GeV, the highest statistics data sample is available from the Jefferson Lab Hall~B CLAS experiment \geleven~\cite{g11}.
Even in this case, some bins in the multidimensional space are unpopulated or subject to large statistical fluctuations. 
This results in large uncertainties in extracting the underlying reaction mechanisms.

The problem has been addressed by studying one or two variables at a time, while integrating over the others.
During integration, correlations between variables, which in turn contain relevant physics information, are partially lost, making the results 
strongly model dependent.
In this context, generative models based on machine learning (ML), which learn the original data distribution and create new so-called {\it synthetic} data that mimic the original distribution, can provide new opportunities for extracting the physics information preserving correlations.
Furthermore, these models can provide another way to extract the `true' values from experimental data removing detector effects, with a procedure known as {\it unfolding}.

Recently, an event-level unfolding analysis using generative adversarial networks (GANs) in inclusive electroproduction was performed~\cite{Alanazi:2020jod}.
The analysis was able to reconstruct accurately single-variable cross sections.
Here, we extend our analysis framework to a multiparticle final state,  
demonstrating for the first time that GANs can be used to reproduce scattering reactions in a higher dimensional phase space. 
Specifically, we optimize our ML analysis framework to the case of two-pion photoproduction at CLAS \geleven kinematics.

This study serves as an excellent testing ground for evaluating the effectiveness of the ML analysis framework in a highly nontrivial case.
The presence of baryon and meson resonances with diverse production mechanisms, which overlap within a limited phase space, generate intricate structures and correlations. 
Moreover, the CLAS detector's highly non-uniform response introduces additional complexities and distortions, adding another layer of complication to the analysis.
To test and validate the framework, we generate Monte Carlo (MC) pseudodata with a realistic model of two-pion photoproduction. 
We produce a synthetic copy with an ``unfolding'' GAN trained on pseudodata that incorporate detector effects  through GEANT simulations~\cite{Brun:1987ma}. 
This would be equivalent to train the GAN with experimental data. 
The detector effects are unfolded using a ``detector-simulation'' GAN, independently trained on a second MC pseudodata sample generated according to phase space and passed through the GEANT model of the detector.
We test the quality of the procedure by a quantitative comparison between the generated MC data and its synthetic copy. 
This closure test, based on MC pseudodata, is a necessary step before applying our analysis framework to experimental data.

The paper is organized as follows: in Sec.~\ref{sec:2pi_photo} we review the importance of two-pion photoproduction in hadron spectroscopy and provide a detailed description of the \geleven kinematics. 
In Sec.~\ref{sec:simulations} we describe the MC framework used to generate pseudodata and incorporate the CLAS detector response. 
In Sec.~\ref{sec:ai_applications} we present the ML framework used for reproducing the detector effects and unfold the `true' distributions from the reconstructed pseudodata. 
The GAN results are reported in Sec.~\ref{sec:closure}, where we compare the generated events with the synthetic copy. 
Finally, in Sec.~\ref{sec:conclusions} we summarize the procedure and outline work in progress to extend the current framework to the analysis of real CLAS data from Jefferson Lab.

\section{Two-pion photoproduction} \label{sec:2pi_photo}

\subsection{The physics case}

The $\pi\pi N$ final state is one of the largest contributors to the total photoproduction cross section off protons at center-of-mass (CM) energies $W\lesssim 2.5$~GeV.
Studies of this final state have considerably extended the available information on the spectrum of the excited states of the nucleon ($N^*$) and their photoexcitation amplitudes.
The quantum numbers of these resonances can be assessed by studying the correlations between the invariant mass and the angular dependencies of their decay products. 
Theoretical estimates based on phenomenological 
approaches~\cite{Capstick:2000qj, Klempt:2009pi, Giannini:2015zia, Qin:2019hgk, Chen:2019fzn}, continuum Schwinger methods \cite{Qin:2019hgk, Chen:2019fzn} as well as from first principles within lattice QCD calculations~\cite{Edwards:2011jj},
have predicted more states than apparently observed in experiments (for reviews, see~Refs.~\cite{Crede:2013kia, Ireland:2019uwn, Beck:2017wkb, Burkert:2022adb}), which is referred to as the `missing baryons' problem.

A strategy to improve the sensitivity to the most elusive states is to impose consistency constraints by performing combined analyses of several final states at once, with $\pi\pi N$ playing a pivotal role for the resonances heavier than 1.6~GeV.
This allows one to disentangle process-dependent nonresonant contributions, and extract the resonance properties in a nearly model-independent manner~\cite{Mokeev:2022xfo}. 
Furthermore, combining photoproduction and electroproduction data has recently proven to be effective in identifying overlapping resonances with the same quantum numbers, as in the case of the $N(1720)$ and $N'(1720)$ states~\cite{CLAS:2002xbv, CLAS:2018drk, Mokeev:2020hhu}.

In the same reaction, by looking at the invariant mass distribution of the $\pi\pi$ pair, one can study  meson resonances, such as the $\rho$ or the $f_2(1270)$. 
While the properties of these resonances are well known, a detailed understanding of their production mechanisms is still missing. 
At low $W \lesssim 2$~GeV one can study how each $N^*$ state contributes to the meson production process.
At higher energies, above the $N^*$ resonance region, the reaction is well described in terms of Regge theory~\cite{JPAC:2016lnm, Mathieu:2018mjw}.
The two energy regimes are smoothly connected, making it nontrivial to study the intermediate region rigorously.
A formalism to do so has been proposed recently for the production of single $\pi$ or $\eta$ mesons~\cite{Mathieu:2020zpm, Mathieu:2018xyc}.
The extension to two-pseudoscalar final states requires having the full multidimensional dependence under control~\cite{JPAC:2021rxu}.
In particular, a complete understanding of meson production mechanisms in the $\pi\pi N$ final state, where resonances are well known, is necessary before facing the more complicated $\eta \pi N$ and $\eta^\prime\pi N$ channels, where exotic hadrons are expected to appear~\cite{Pauli:2021gde}.

\subsection{$\gamma p \to \pi^+ \pi^-p$ kinematics}\label{sec:2pi_photo-kin}

Measurement of the three-body final state in two-pion photoproduction represents a significant challenge to experiment.
Recently a large body of data 
on $\pi^+\pi^-p$ photoproduction observables has become available from measurements by the CLAS Collaboration, with $W \le 2.9$~GeV~\cite{Ireland:2019uwn, CLAS:2018drk, CLAS:2009ngd, CLAS:2005oqk, CLAS:2001zxv}. 
For a given collision energy, the differential cross section for this process depends on five independent variables, which can be chosen to be the invariant masses of the two pions, $M_{\pi^+\pi^-}$, and the proton-$\pi^-$ pair, $M_{p\pi^-}$, and three angles in the CM frame.
Two of the angles are the polar angle $\theta_{\pi^+}$, with the $z$-axis along the photon three-momentum, and the angle $\alpha_{[\pi^+ p][\pi^-p']}$ between the plane containing the initial target proton $p$ and $\pi^+$ three-momenta and the plane containing the $\pi^-$ and recoiling proton $p'$ three-momenta. 
An equivalent choice would replace $\theta_{\pi^+}$ with the invariant momentum transferred $t_{\pi^+}$, defined as the difference squared between the photon and $\pi^+$ four-momenta. 
The fifth variable $\phi$ is the azimuthal angle of $\pi^-$ with respect to the plane containing the photon three-momentum and the polarization vector, and is relevant only in experiments with polarized beam or target. For unpolarized data, one can still define $\phi$ by pointing the polarization vector in an arbitrary direction, resulting in a $\phi$-independent cross section.
Other possible choices for variables are $M_{p\pi^+}$ (invariant mass of the proton-$\pi^+$ pair), $t_{\pi^-}$ (momentum transferred between photon and $\pi^-$), $t$ (momentum transferred between target and recoil protons), or $\cos\theta$ (cosine of the angle between target and recoil protons in the CM frame).

\begin{figure}[t]
\begin{center}
\includegraphics[width=0.95\columnwidth]{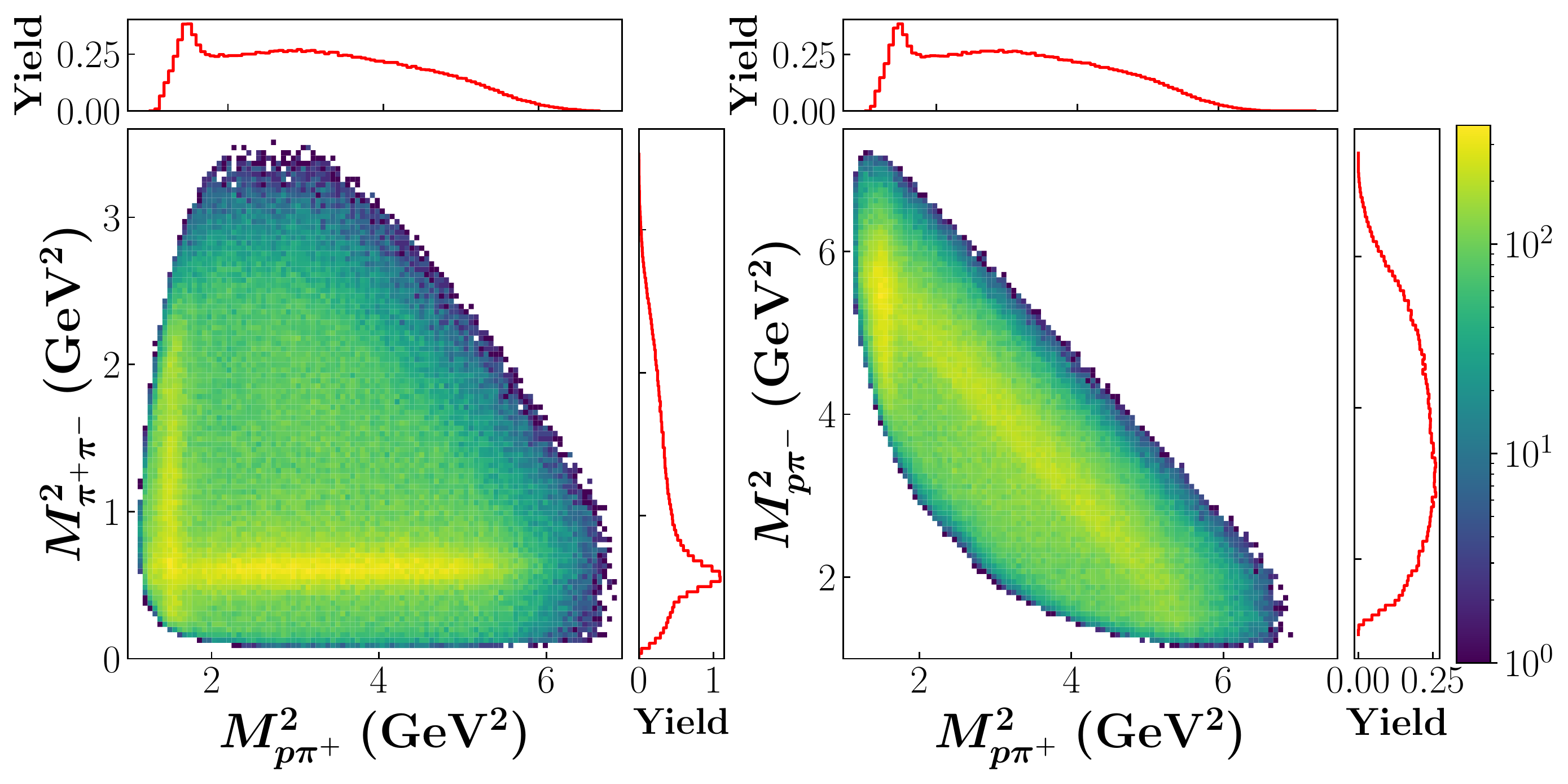}

\vspace{-0.3cm}
\caption {Examples of 2D normalized yield distributions and their 1D projections measured in CLAS \geleven experiment (before unfolding detector effects)~\cite{CLAS:2009ngd}. Plots show invariant mass distributions for the yields $M^2_{\pi^-\pi^+}$ versus $M^2_{p \pi^+ }$ (left) and $M^2_{p \pi^-}$ versus $M^2_{p \pi^+}$ (right). For each 2D plot, the correspondent 1D projections are shown on top and right. The distributions are dominated by the prominent contributions from $\Delta^{++}$ and $\rho$ resonances, which appear as bands in the 2D distributions and peaks in the appropriate 1D projection. In particular, the $\Delta^{++}$ appears as a vertical band in the 2D plots, and as a peak in the 1D $M^2_{p \pi^+}$ projection. The $\rho$ is the horizontal band in the left 2D plot, the diagonal band in the right 2D plot, and the peak in the 1D $M^2_{\pi^+\pi^-}$ yields.  From the right plots one can see that, choosing $M^2_{p \pi^+}$ and  $M^2_{p \pi^-}$ as independent variables, the existence of the $\rho$ is lost if one studies the 1D projections only.}
\label{1d2d_yeilds}
\end{center}
\end{figure}

Multidimensional analyses are becoming standard, albeit computationally difficult, in modern high statistics experiments~\cite{LHCb:2021uow, LHCb:2022jad}.
However, some specific reactions can suffer from limited statistics.
In particular, the direct extraction of $\pi^+ \pi^- p$ photoproduction events at a given $W$ value, on a 5D grid (or 4D, if integrated over the angle $\phi$) with a bin size acceptable for physics analyses, is quite challenging.
Even the highest statistics $\pi^+ \pi^- p$ photoproduction sample collected with CLAS~\cite{CLAS:2018drk, CLAS:2009ngd} results in a limited number of counts in the 4D cells (typically $<10$ events per cell).
In Ref.~\cite{CLAS:2018drk}, theoretical curves were fitted to the marginal 1D distributions, determined by integrating the acceptance- and efficiency-corrected 5D distribution over the remaining four variables.
This procedure largely washes out the correlations present in the original data, leading to a significant loss of relevant information contained in the joint distribution. 
In this paper we aim to overcome this problem with ML techniques.

To illustrate this, in Fig.~\ref{1d2d_yeilds} we show two examples of 2D~distributions and their 1D~projections, as measured in CLAS \geleven experiment without efficiency corrections~\cite{CLAS:2009ngd}. 
From these distributions one immediately sees the presence of intermediate resonances that appear as enhancements in the invariant mass of the system in which they decay.
For example, the band at $M^2_{p \pi^+}\simeq 1.5\text{~GeV}^{2}$ corresponds to the $\Delta(1232)$ baryon resonance, which appears as an intermediate unstable state in the reaction $\gamma p \to \Delta^{++}\pi^- \to p \pi^+ \pi^-$. 
The band centered at $M^2_{\pi^+ \pi^-} \simeq 0.6\text{~GeV}^2$ corresponds instead to the $\rho(770)$ meson resonance, in the reaction $\gamma p \to p\,\rho^0 \to p \pi^+\pi^- $. 
The two resonances are clearly visible as bumps in the respective 1D projections.
Looking at 1D projections only, one can easily miss the presence of a resonance if the relevant invariant mass distribution is not explicitly considered.
This is an example of loss of information that is contained in correlations.
Moreover, 
because of quantum interference, the production of $\rho^0$ and $\Delta^{++}$ are not independent processes, and it is impossible to associate one event exclusively with either process.
This interference appears in the correlations between the invariant masses, and can be partially lost in the 1D projections.

\subsection{Two-pion photoproduction with CLAS} \label{sec:clas}

The CLAS spectrometer in Hall~B at Jefferson Lab was based on a $\sim 1.25$~T toroidal magnet which bends charged particles produced in the hadronic interaction along the polar angles $\theta_\text{lab}$ (the $z$-axis along the photon beam), while the preserving azimuthal angles $\phi_\text{lab}$.
The polarity of the field determined if positive/negative charges were bent towards/away from the beam line into the acceptance of the detector.
A system of three layers of multi-wires drift chambers~\cite{Mestayer:2000we} provided momentum information with the resolution, $\sigma_p/p$, ranging from 0.5 to 1.0\%, depending on the kinematics.
Charged hadron identification was obtained by time-of-flight scintillators~\cite{Smith:1999ii}.
Photoproduction experiments were conducted with a bremsstrahlung photon beam produced by the CEBAF continuous electron beam impinging on $8 \times 10^{-5}$ radiation lengths  thickness gold foil.
A bremsstrahlung tagging system~\cite{Sober:2000we} with a photon energy resolution of $0.1\%$ was used to measure the photon energy in each recorded event.
The target cell was a 4~cm in diameter and 40~cm long Mylar cylinder, filled with liquid hydrogen at 20.4~K.

The experimental conditions reported in this paper, and simulated in the framework described in Sec.~\ref{sec:simulations}, correspond to the \geleven experiment that ran in CLAS in 2004.
During the experiment, the torus field was such that positive particles were bent away from the beam line.
The detector geometrical acceptance for each positive particle in the relevant kinematic region was about 40\% and somewhat less for negative particles (bent towards the beamline and out of the detector acceptance). 
The primary electron beam energy was 4.02~GeV, providing a tagged photon beam in the energy range from 0.8 to 3.8~GeV. 
For this analysis we focus on the highest energy region, 3.0--3.8~GeV, that was analyzed in~Ref.~\cite{CLAS:2009ngd}.

The exclusive reaction $\gamma p \to \pi^+ \pi^-p$  was isolated by detecting the  proton and the $\pi^+$ in the CLAS spectrometer, while the  $\pi^-$ was reconstructed from detected particle four-momenta using the missing-mass technique.
In this way, the exclusivity of the reaction was ensured, keeping the contamination from the multipion background to a minimum level.
Only events within a fiducial volume were retained in the analysis, in order to avoid the regions at the edge of the detector acceptance. 
Cuts were defined on the minimum proton momentum and the hadron minimum and maximum polar angle.
After all the cuts, approximately 40~M events were identified as produced in exclusive two-pion photoproduction, making the \geleven dataset the largest statistics sample of this reaction in the above photon energy range. 
Details of the \geleven analysis can be found in Ref.~\cite{CLAS:2009ngd}.

\begin{figure}[t]
\begin{center}
\includegraphics[width=0.95\columnwidth]{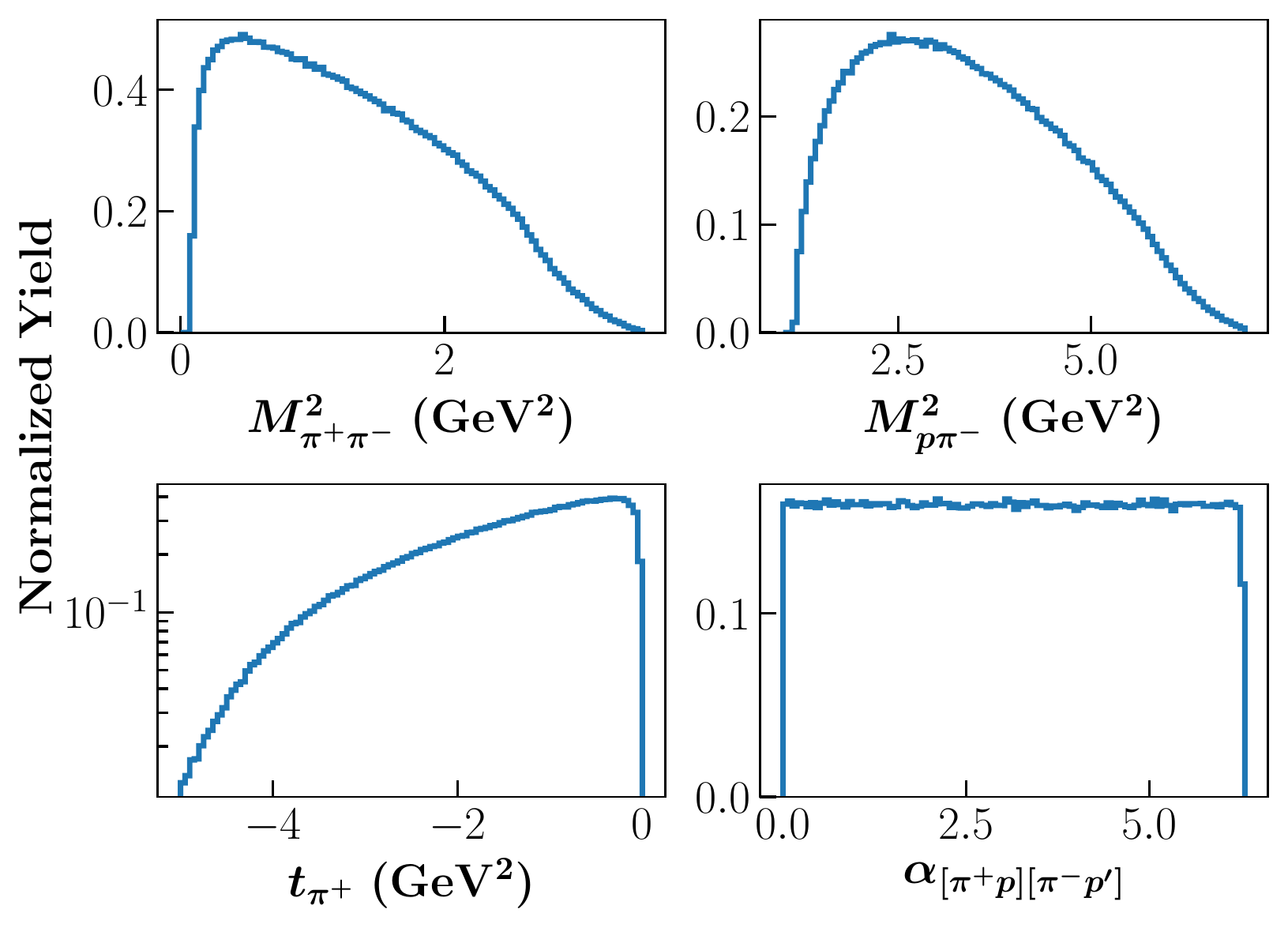}
\vspace{-0.3cm}
\caption{Normalized 1D projections of events generated with PS-MC:  $\pi^+\pi^-$ invariant mass squared (top-left), $p\pi^-$ invariant mass squared (top right), square of the four-momentum transferred from the target to the recoil proton, $t_{\pi^+}$ (bottom-left), the angle $\alpha_{[\pi^+ p][\pi^-p']}$ in the CM frame between the plane containing the initial target $p$ and $\pi^-$ three-momenta and the plane formed by $\pi^+$ and the scattered proton three-momenta (bottom right).} 
    \label{fig:PS-1d-distributions}
\end{center}
\end{figure}

\section{MC simulation frameworks}\label{sec:simulations}

In this section we describe the simulation frameworks used to perform the closure test.
Pseudodata corresponding to two-pion photoproduction in the kinematics of the \geleven experiment were generated using two different MC event generators that produce the four-momenta of the final state particles.  
A realistic GEANT simulation was used to reproduce the finite resolution and limited acceptance of the CLAS detector.

Detector effects were assessed with a first MC generator based on a pure phase-space distribution. 
To perform the closure test, we deployed a second MC generator based on a realistic physics model. 
The use of two different MC generators minimizes the model dependence in the extraction of the original information and mimics a real situation, where the detector effects are estimated with simulations that are similar but not identical to the experimental distributions.

\subsection{Two-pion event generators}
\begin{figure}[t]
\begin{center}
\includegraphics[width=0.95\columnwidth]{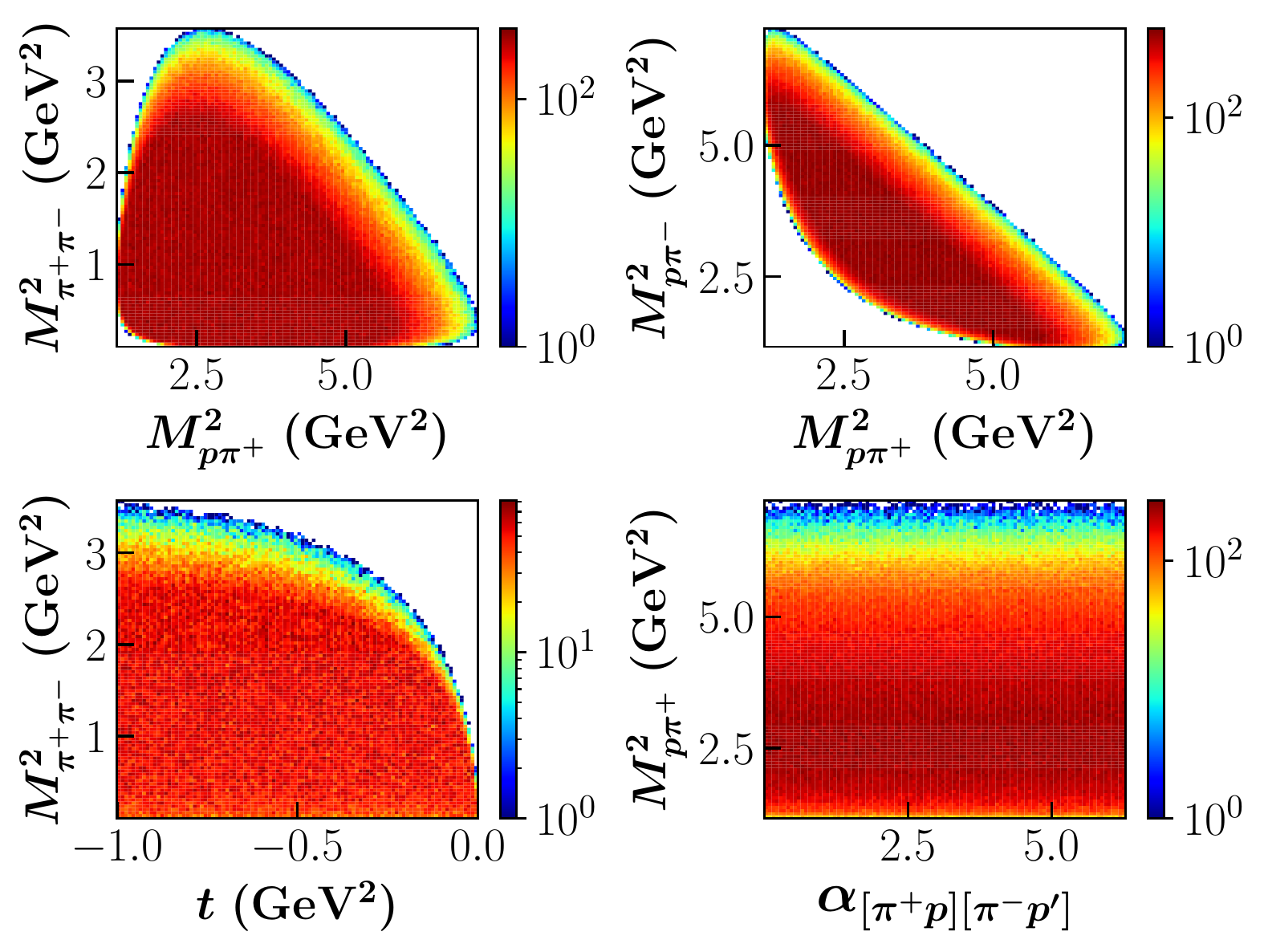}
\vspace{-0.3cm}
\caption{Normalized 2D distributions generated with PS-MC: invariant mass distributions $M^2_{\pi^+\pi^-}$ versus $M^2_{p\pi^+}$ (top left), and $M^2_{p\pi^-}$ versus $M^2_{p\pi^+}$ (top right), $M^2_{\pi^+\pi^-}$ invariant mass versus momentum transfer $t$ (bottom left), and CM angle versus $M^2_{p\pi^+}$ (bottom right).}

    \label{fig:PS-2d-distributions}
\end{center}
\end{figure}

The two MC generators simulate the interaction of an incoming unpolarized photon beam with a bremsstrahlung spectrum, in the energy range $3.0$--$3.8$~GeV, with a target proton at rest. 
With the choice of variables described in Sec.~\ref{sec:2pi_photo-kin}, the yields are proportional to the differential cross section, and thus to the squared of the production amplitude $A$ summed over polarizations, 
\begin{align}
&\frac{\dd^5 \sigma}{\dd{M}^2_{p\pi^-}\dd{M}^2_{\pi^+\pi^-}\dd{t}_{\pi^+}  \dd\alpha_{[\pi^+ p][\pi^-p']}\dd\phi} 
\nonumber\\
&\propto \Big[\left(W^2 - (M_{p\pi^-} + m_\pi)^2\right)\left(W^2 - (M_{p\pi^-} - m_\pi)^2\right)\Big]^{-1/2}
\nonumber\\
&\times\sum_\text{pol}\left|A \big(M^2_{p\pi^+},M^2_{\pi^+\pi^-},\cos \theta_{\pi^-},\alpha_{[\pi^+ p][\pi^-p']}\big)\right|^2~.
\end{align}
The first MC generator, referred to as {\it phase space} or PS-MC, distributes final state events according to the $\pi^+ \pi^-p$ phase space. 
This corresponds to assuming that the production amplitude is a constant.
This is clearly unrealistic since, as discussed above, two-pion photoproduction has a much more complicated structure. 
However, it has the advantage of being well-defined, agnostic to physics models, and distributes events uniformly across the full reaction kinematics.
The 1D-projected PS-MC event distributions are shown in Fig.~\ref{fig:PS-1d-distributions}, while the 2D distributions are illustrated in Fig.~\ref{fig:PS-2d-distributions}.

\begin{figure}[t]
\begin{center}
\includegraphics[width=0.95\columnwidth]{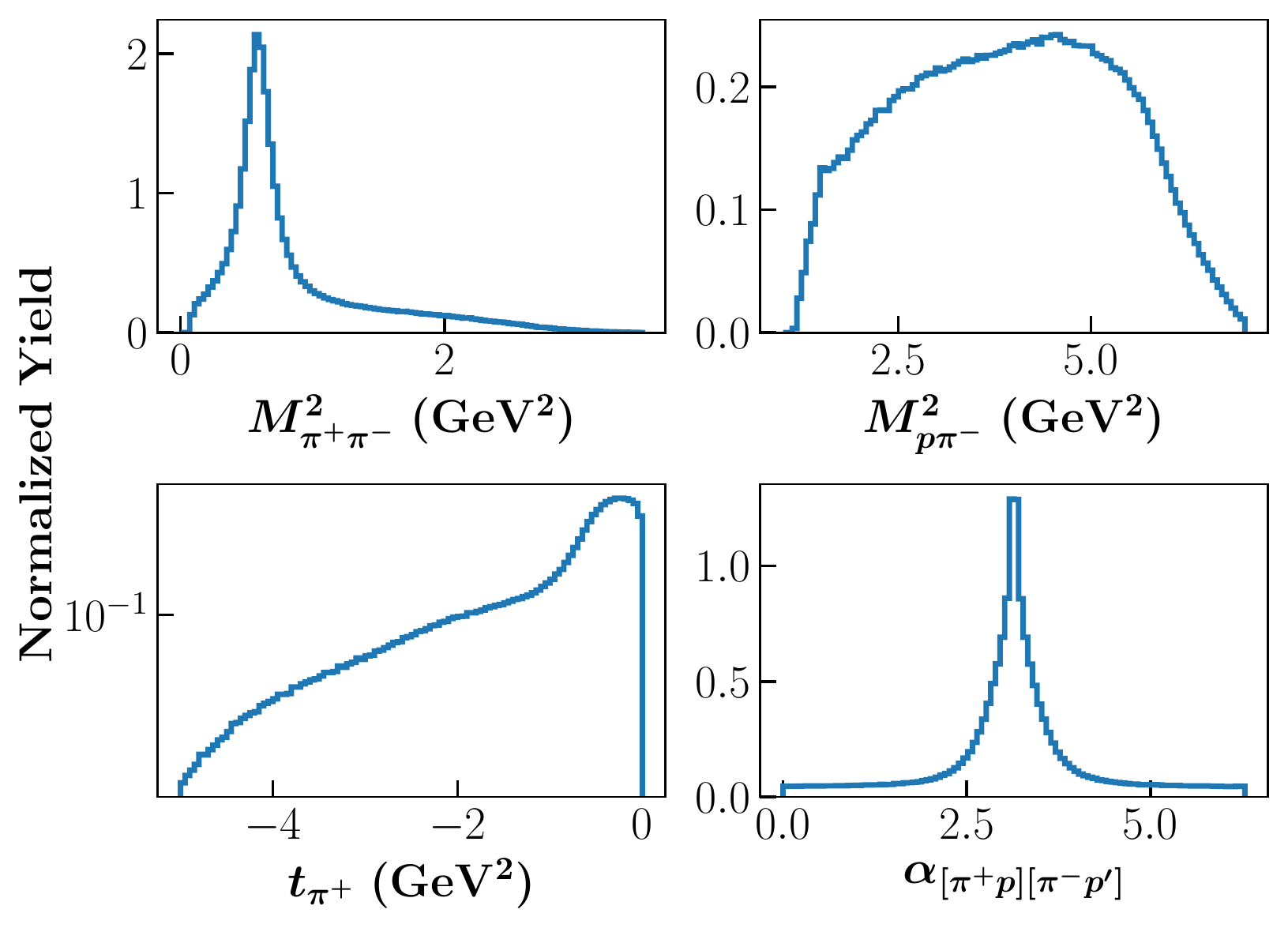}
\vspace{-0.3cm}
\caption{Normalized 1D projections of events generated with RE-MC. The panel descriptions are as in Fig.~\ref{fig:PS-1d-distributions}.} 
    \label{fig:RE-1d-distributions}
\end{center}
\end{figure}

\begin{figure}[t]
\begin{center}
\includegraphics[width=0.95\columnwidth]{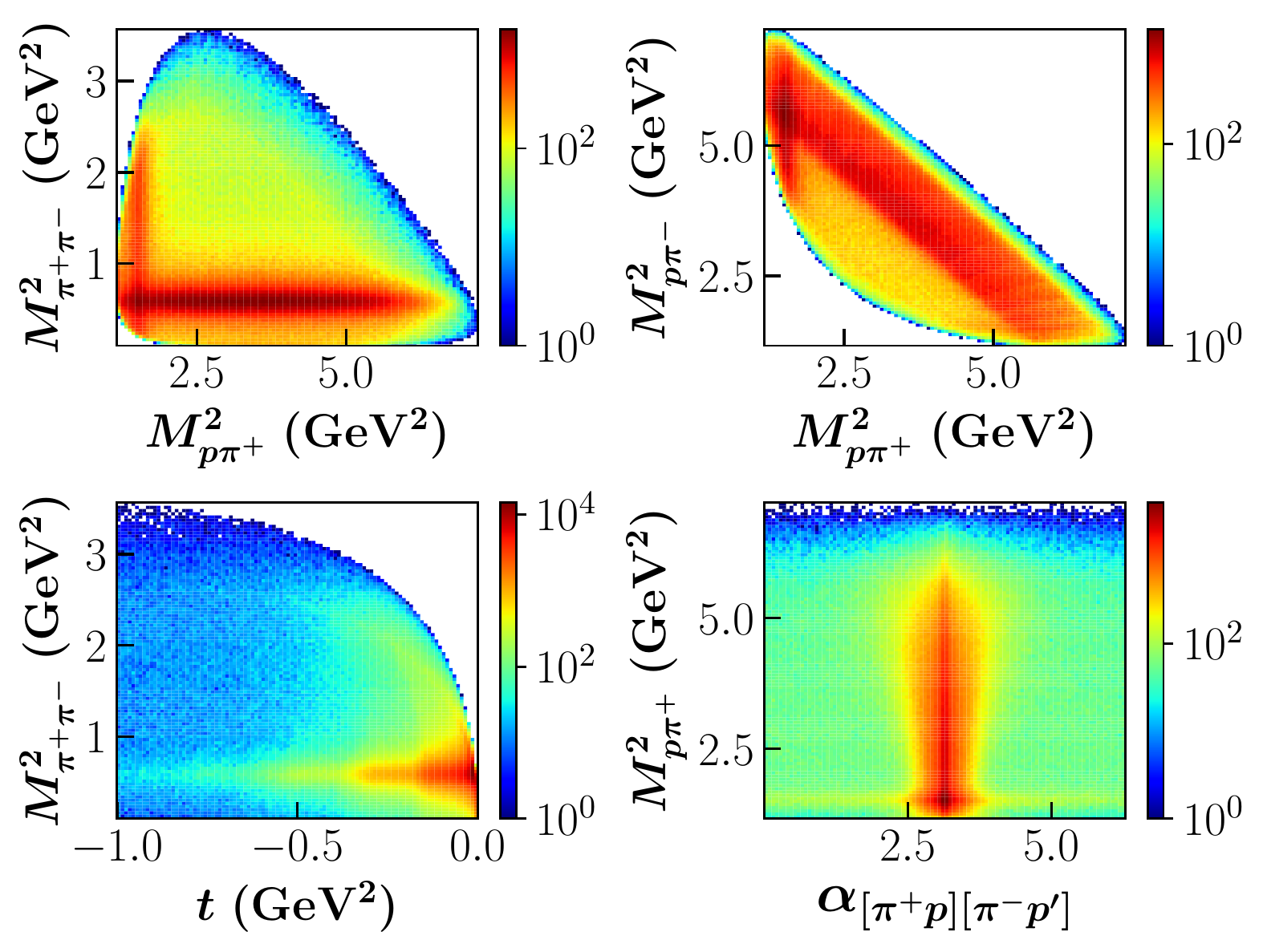}
\vspace{-0.3cm}
\caption{2D distributions generated with RE-MC. The panel descriptions are as in Fig.~\ref{fig:PS-2d-distributions}.}
    \label{fig:RE-2d-distributions}
\end{center}
\end{figure}

The second MC event generator, which we refer to as {\it realistic} or RE-MC, considers the amplitude squared as an incoherent sum of the three dominant intermediate resonances observed,
$\gamma p \to \left(p \rho^{0},\,\Delta^{++} \pi^- ,\,\Delta^{0} \pi^+ \right)\to \pi^+ \pi^-p$,
added to a $\sim 10\%$ constant that mimics the nonresonant two-pion photoproduction contribution.
Each process has been weighted with the corresponding contribution to the total cross section as reported in Ref.~\cite{Whalley:1989mt}. 
The angular distributions relative to resonance production are parametrized from measured differential cross sections reported in the same database.
The decays $\rho\to \pi\pi$ and $\Delta\to p \pi$ are described using the correct spin structure with the decay matrix elements detailed in Ref.~\cite{Schilling:1969um}.
The resulting 1D and 2D projections for events generated by RE-MC are shown in Figs.~\ref{fig:RE-1d-distributions} and~\ref{fig:RE-2d-distributions}, respectively.
We note that this model neglects the interference terms between the intermediate resonances.
Despite this, the resulting distribution provides a reasonable description of the experimental data, showing resonance structures in the invariant masses and the correct angular behavior of particles in the final states.

\subsection{CLAS detector simulation}

The CLAS detector response has been simulated using the standard GEANT Monte Carlo simulation package, GSIM, used by the CLAS Collaboration~\cite{GSIM}.
It consists of a central steering and control package that calls a number of independent detector geometry and response packages. 
A post-processing code (GSIM-Post-Processor or GPP) has been used to fine tune the GSIM output to match the tails of the experimental resolution and other effects, such as the detector's dead channels, not described by the idealized GEANT-based simulation.
The GSIM output has been fed to the same reconstruction code, RECSIS, used to process experimental data.  
We will refer to REC or {\it detector-level} events to identify the set of pseudodata as processed by the detector simulation, while GEN or {\it vertex-level} will identify the `true' events as generated by the MC code.

\begin{figure}[t]
\begin{center}
\includegraphics[width=0.95\columnwidth]{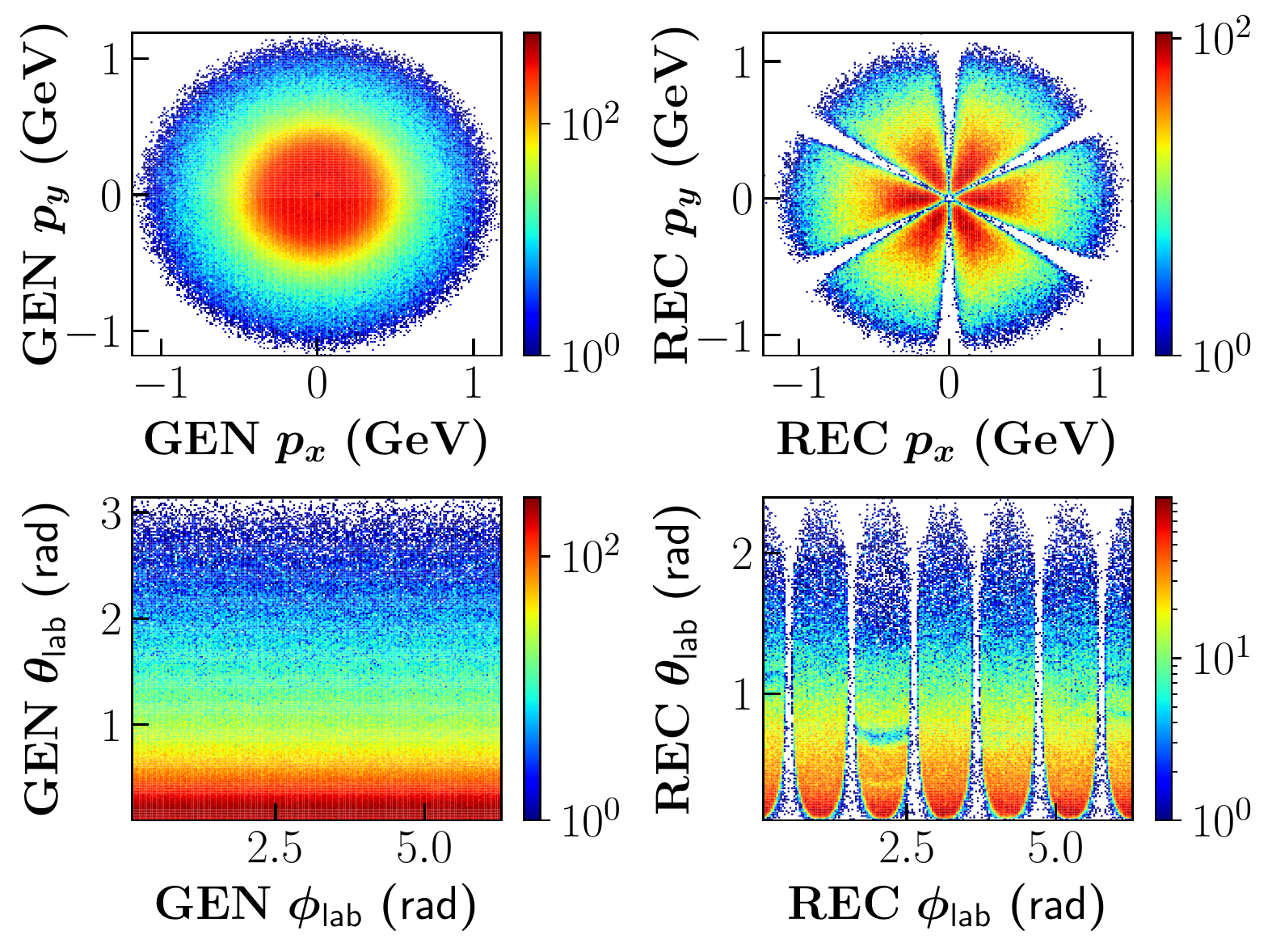}
\vspace{-0.3cm}
\caption{The $\pi^+$ kinematic variables in the laboratory reference frame as GENerated with RE-MC (left) and REConstructed by CLAS (right): $p_x$ versus $p_y$ (top panels) and $\theta_\text{lab}$ versus $\phi_\text{lab}$ (bottom panels).}
    \label{fig:GSIM-acc}
\end{center}
\end{figure}

As reported in Sec.~\ref{sec:clas}, the CLAS detector has a nonuniform acceptance, reduced in the azimuthal angle $\phi_\text{lab}$ (around the beam) by the presence of the six coils of the toroidal magnet, and in the polar angle $\theta_\text{lab}$ (with respect to the beam direction) by the limited area covered by the drift chambers, calorimeter and time-of-flight systems.
A further limitation concerns the minimum accepted momenta of charged hadrons, due to the energy loss in materials crossed along the track and to the effect of the toroidal magnetic field that bends low-momentum particles out of the detector acceptance.
The limited CLAS acceptance results in a reduced yield in REC with respect to  GEN events, since not all generated events are reconstructed.
The effect of the CLAS acceptance on the $\pi^+$ variables in the laboratory frame is shown in Fig.~\ref{fig:GSIM-acc}.

\begin{figure}[t]
\begin{center}
\includegraphics[width=0.95\columnwidth]{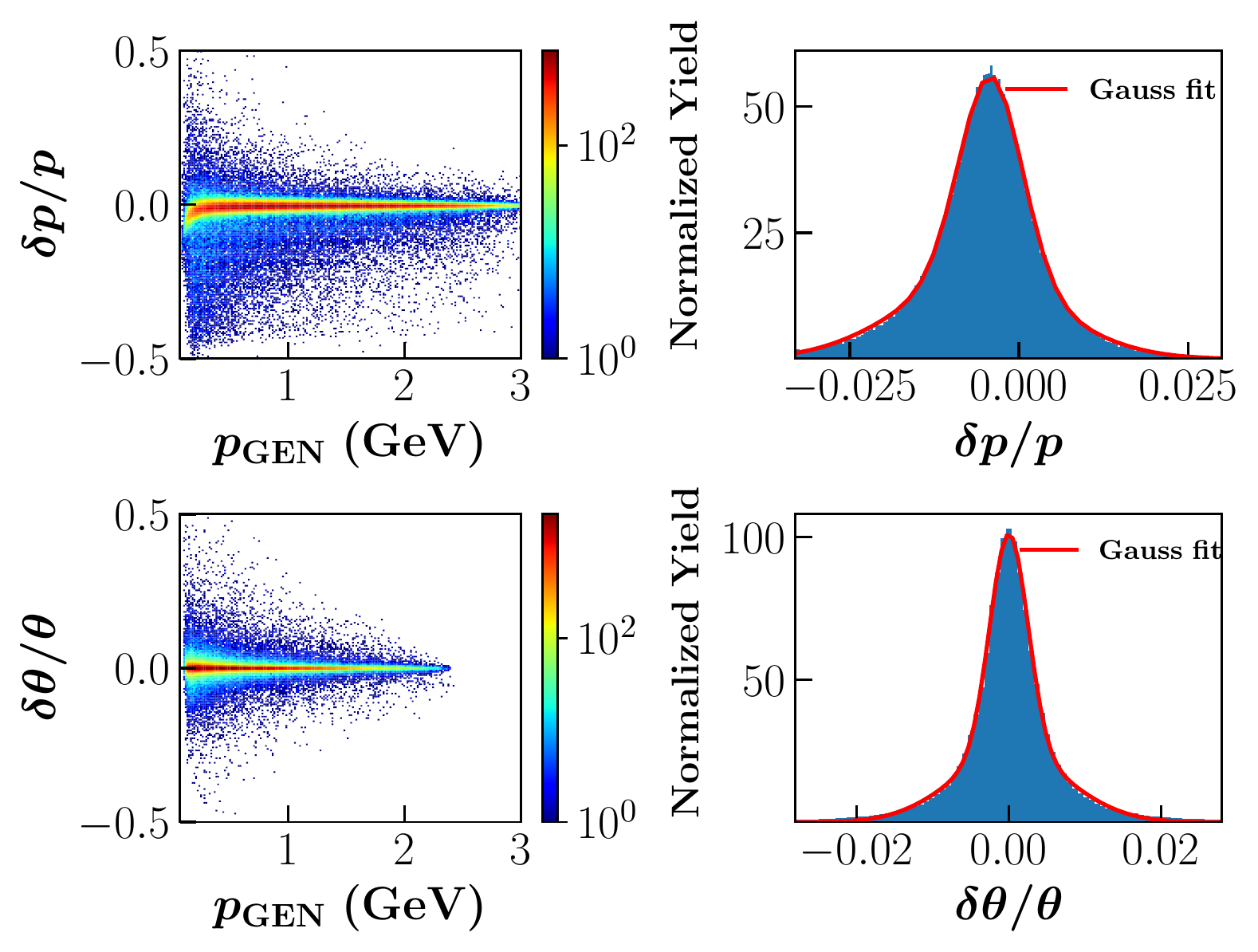}
\vspace{-0.3cm}
\caption{The smearing of $\pi^+$ kinematic variables in the laboratory reference frame. Top: relative momentum resolution $(p_\text{REC}-p_\text{GEN})/p_\text{GEN}$ as a function of $p_\text{GEN}$. Bottom: relative angular resolution defined as $(\theta_\text{REC}-theta_\text{GEN})/\theta_\text{GEN}$ as a function of $p_\text{GEN}$. The right panels display the 1D projections of the 2D distributions as histograms, and the red lines represent double Gaussian fits. Pseudodata events were generated with RE-MC.}
\label{fig:GSIM-res}
\end{center}
\end{figure}

As any detector, CLAS has finite resolution, which `smears' the measured kinematic variables resulting in a difference between REC and GEN, even when the event is accepted.
The smearing affects the reconstructed three-momenta of any detected particle within the CLAS acceptance with a distortion depending on the three-momentum of the particle.
Figure~\ref{fig:GSIM-res} shows the resolution on the detected (REC) $\pi^+$ momentum and polar angle as a function of the `true' (GEN) momentum, along with the projections in 1D corresponding to the CLAS relative momentum and angular resolution.
Fitting the two curves to a double Gaussian line, we obtained $\delta p / p \sim 0.8\%$ and $\delta \theta / \theta \sim 0.5\%$. A similar smearing affects the kinematic variables of the detected proton.

The resolution of the CLAS detector is sufficiently high so as to allow the use of the missing mass technique to identify the exclusive two-pion reaction against the multipion background.
The technique uses knowledge of the initial state and of the detected particles to calculate the invariant mass of the undetected system to fulfill energy-momentum conservation, within detector resolution.
If all particles are detected, the missing mass is zero. 
If a single particle is undetected, its mass appears as a peak in the missing mass spectrum.
If two or more particles are lost, the missing mass of the system is unconstrained and does not peak, but rather distributes smoothly. 
The technique is only applicable if the experimental resolution is sufficient to disentangle the missing mass peak from this multiparticle background.
Clearly, the more particles are detected, the lower is the resolution for the missing mass due the error propagation, limiting the validity of the technique to reactions with a small number of particles in the final state.
When the missing particle has been identified, its four-momentum is determined by energy and momentum conservation, and the final state can be fully reconstructed.

In two-pion photoproduction, the requirement of at most a single undetected particle corresponds to the following topologies (missing particle in parentheses): $p\pi^+(\pi^-)$, $p\pi^-(\pi^+)$, $\pi^+\pi^-(p)$ and $\pi^+\pi^-p$ (all three detected). 
Considering the CLAS acceptance, the yield of different topologies is quite different, with a ratio of $(100 :37:30:35)$ for the respective topologies.
Since the $p\pi^+(\pi^-)$ is by far the dominant contribution to REC data, we focus on this topology, although similar conclusions also hold for the others.
Each topology is in one-to-one correspondence with different areas of the allowed phase space, and a combination of different topologies would therefore extend the kinematic coverage of the measurement, mitigating the effect of the limited detector acceptance.

\begin{figure}[b]
\begin{center}
\includegraphics[width=0.75\columnwidth]{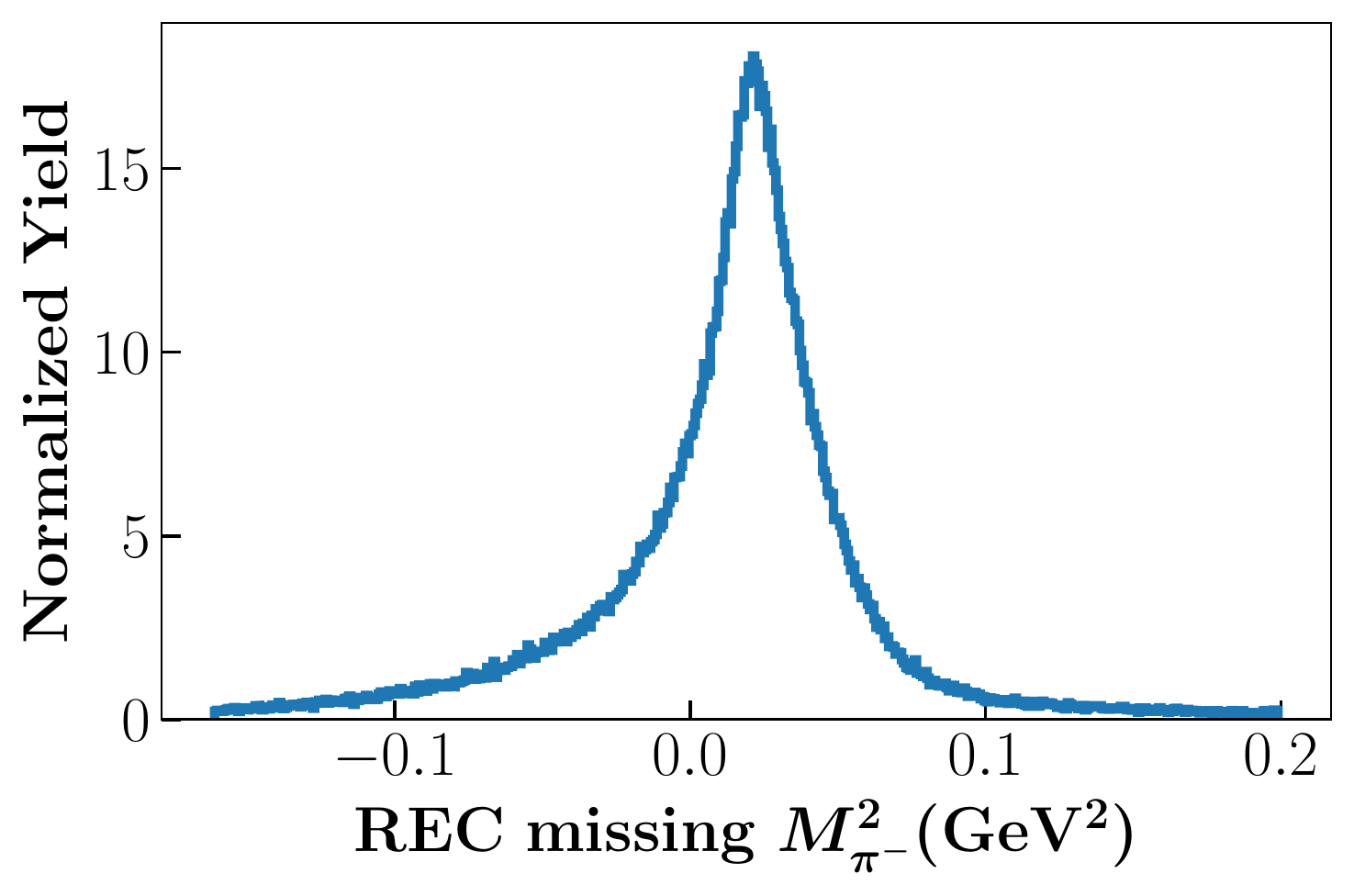}
\vspace{-0.3cm}
\caption{The $\pi^-$ missing mass spectrum for the $p\pi^+(\pi^-)$ topology from REC data. Pseudodata events were generated with RE-MC.}
\label{fig:GSIM-MissMass}
\end{center}
\end{figure}

Figure~\ref{fig:GSIM-MissMass} shows the missing mass distribution of the $p\pi^+(\pi^-)$ topology. 
This exclusive final state is identified by selecting events with missing mass in the peak.
Since these simulations only contain the two-pion final state, no multiparticle background populates the plot. 
The equivalent distribution for \geleven data shows a significant multipion background~\cite{CLAS:2009ngd} that populates the positive side of the missing mass spectrum, and is rejected during the analysis to assure the reaction exclusivity.

\begin{figure*}[t]
\centering
\vspace*{-30mm}
\includegraphics[width=1\textwidth]{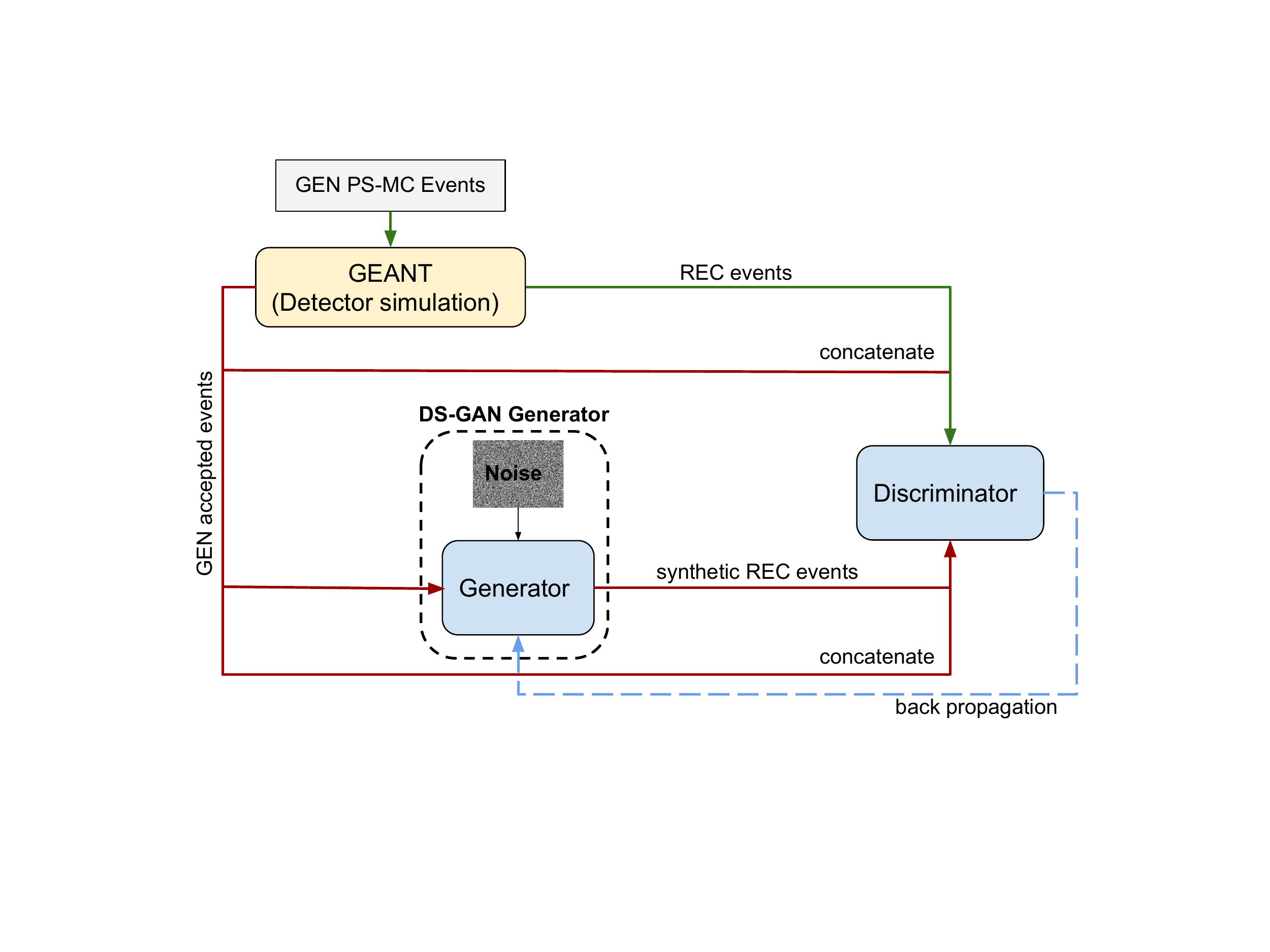}
\vspace*{-39mm}
\caption{Schematic view of the ML detector simulation GAN (DS-GAN), where the GAN generator converts input GEN vertex-level events features and noise to
REC detector-level events. The training is performed on PS-MC pseudodata passed through the GEANT simulation. Synthetic REC and REC pseudodata are concatenated with GEN PS-MC events and fed to the discriminator.
}
\label{fig: Inner GAN}
\end{figure*}

\section{GAN-based unfolding methodology}
\label{sec:ai_applications}

GANs, a type of neural networks that have gained significant attention in recent years, are powerful generative models highly effective in generating high-quality, realistic data in various fields~\cite{goodfellow2020generative}. 
The architecture of a typical GAN involves a generator network that learns to produce data and a discriminator network that learns to differentiate between the generated and reference data. 
The two networks are trained alternately in a competitive setting, where the generator tries to produce more realistic data to fool the discriminator, and the discriminator tries to correctly identify the generated data. 
This iterative process leads to the generation of data that are progressively more realistic, with the ultimate goal of producing synthetic data that are indistinguishable from the reference data.

GANs have been widely applied in many domains, such as image synthesis~\cite{karras2019style}, text generation~\cite{li2018generative}, music composition~\cite{brunner2018symbolic}, and videos~\cite{clark2019adversarial}, and have demonstrated impressive results. In image synthesis, GANs have been used to generate highly realistic images visually indistinguishable from real images, which has numerous practical applications in fields such as gaming, film, and art.

Successfully training GANs can be notoriously challenging, however.
Numerous GAN models experience significant issues, such as mode collapse, non-convergence, model parameter oscillation, destabilization, vanishing gradients, and overfitting, resulting in an unbalanced training of the generator and discriminator~\cite{salimans2016improved, arora2017gans, arjovsky2017towards, bang2021mggan}.
In contrast to typical GAN applications, the success of a GAN-based event generator in nuclear and particle physics depends on its ability to accurately reproduce correlations among the momenta of the particles, which becomes increasingly challenging beyond two dimensions. 
Moreover, the multidimensional momentum distributions of events associated with nuclear and high-energy physics reactions, such as the two-pion photoproduction process considered in this work, exhibit highly complex patterns and range over orders of magnitude across the phase space.
The task of developing an appropriate GAN architecture that is able to simultaneously reproduce all the correlations among particle momenta, and accurately reproduce multidimensional histograms, is therefore rather difficult.

Machine learning event generators have gained prominence as efficient fast simulation tools in various scientific fields, including high-energy and nuclear physics~\cite{gao2020event, otten2021event, paganini2018calogan, ijcai2021p293, hashemi2019lhc, butter2019GAN, DiSipio:2019imz}. 
Unlike traditional simulation methods that rely on a theoretical framework for the underlying reaction, machine learning event generators learn from large datasets and use this knowledge to produce new events with high fidelity. 
GANs have emerged as powerful tools in the field of fast simulation, where they learn to generate events that closely resemble reference data, capturing the underlying physics processes and their distributions~\cite{paganini2018accelerating, de2017learning, musella2018fast, hashemi2019lhc, butter2019GAN}.

Furthermore, GANs have been employed to address the challenge of simulating detector effects in fast simulation~\cite{bellagente2020gan, Alanazi:2020jod}. This application of GANs helps bridge the gap between simulated and reference data, enabling more realistic and precise simulations for experimental analyses.
A comprehensive survey of existing ML-based event generators can be found in Ref.~\cite{ijcai2021p588}.
In this study, we employ the architectural framework of the Least Squares GAN, which involves substituting the cross entropy loss function in the discriminator component of a conventional GAN with a least square term.
For further details, see Ref.~\cite{mao2017squares}.

\begin{figure*}[t]
    \centering
    \includegraphics[width=1\textwidth]{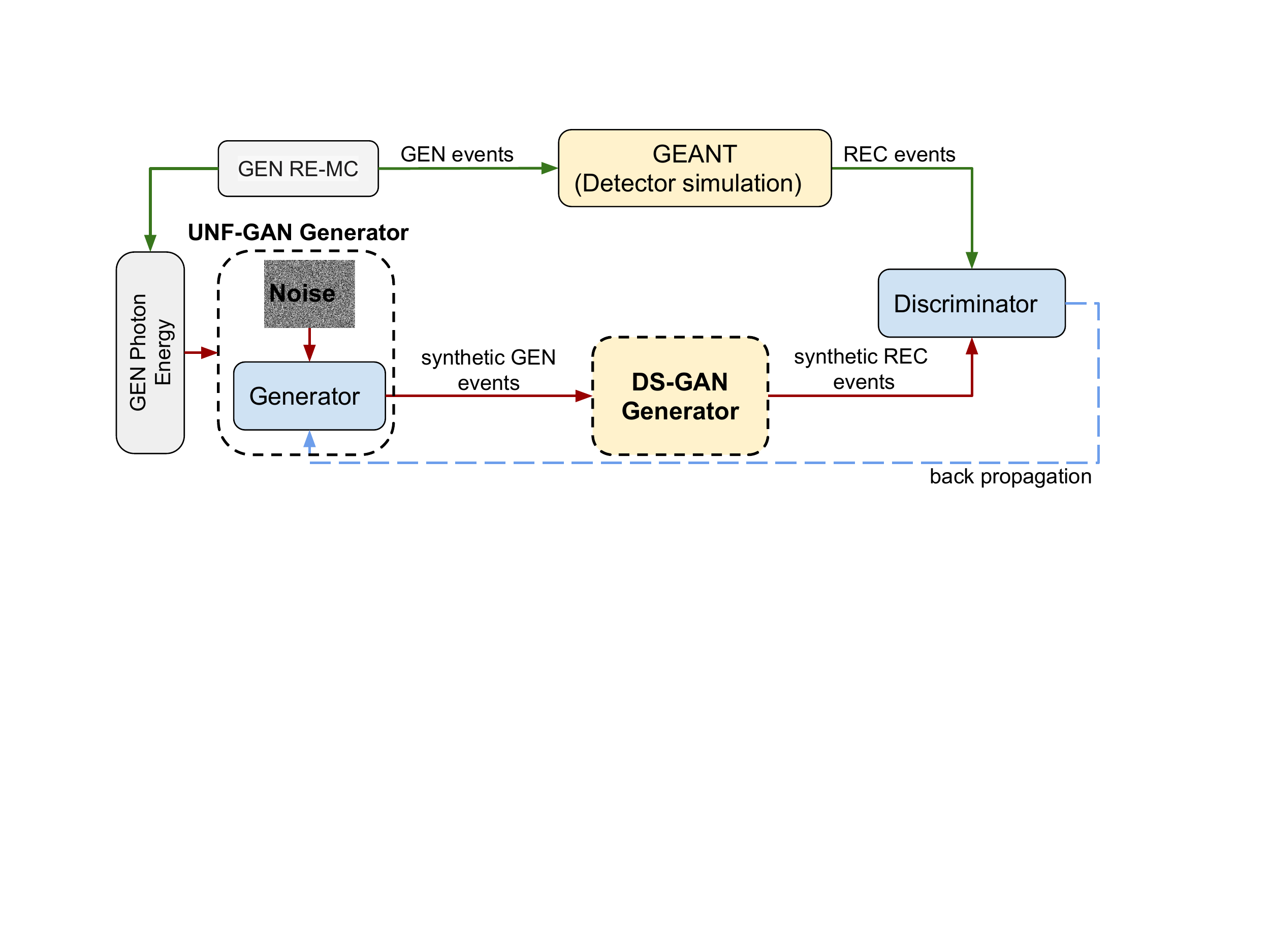}
    \vspace*{-60mm}
    \caption{Schematic view of the UNF-GAN training framework: the UNF-GAN utilizes a generator that converts a GEN photon energy and random noise into synthetic GEN event features, which are passed through the DS-GAN to incorporate detector effect, and are converted into synthetic REC event features.
    The generator is updated using gradients constructed by a deep neural network discriminator, which compares the features of synthetic and reference REC detector-level events obtained through GEANT.} 
    \label{fig: GAN framework}
\end{figure*}

In the following, we describe the GAN architecture used to generate the synthetic data that reproduce the $\gamma p \to \pi^+ \pi^-p$ RE-MC pseudodata.
As mentioned above, two different GANs were developed and combined.
The detector simulation GAN (DS-GAN) was trained on PS-MC pseudodata to learn the detector effects, and was later inserted between the generator and the discriminator of the unfolding GAN (UNF-GAN) to unfold the GEN vertex-level information from REC pseudodata.

\subsection{Detector simulation GAN (DS-GAN)}
\label{sec:DS-GAN}

In order to capture the detector effects, we have developed an ML-based detector simulation using a conditional GAN~\cite{mirza2014conditional}, as illustrated in Fig.~\ref{fig: Inner GAN}. 
Our approach involves training a conditional GAN generator to simulate the detector's smearing effect so that it generates synthetic REC detector-level events from input noise and PS-MC GEN events. 
The GEN PS-MC accepted events are passed through the GEANT chain to obtain REC pseudodata.
As proposed by Bellagente {\it et al.}~\cite{bellagente2020gan}, both the synthetic REC and REC pseudodata are ``concatenated'' with original GEN events and fed to the GAN discriminator as input to facilitate convergence.
After successful training, the DS-GAN generator serves as the ML detector surrogate that will be integrated into the UNF-GAN architecture.

Summarizing the model architecture of the DS-GAN, the generator, conditioned on accepted events (GEN), takes in as input a 100-dimensional array of random values with a mean of 0 and a standard deviation of~1.
The generator network consists of five hidden layers, each with 128 neurons, using a leaky rectified linear unit (ReLU) activation function. The final hidden layer is connected to a four-neuron output layer, which uses a linear function to represent the generated features. 
At the end of the training, the DS-GAN generator learns how to convert the GEN accepted events into REC events, effectively mimicking the smearing due to the detector as described by GEANT.

The discriminator is made of a neural network with five hidden dense layers. The first three layers have 256 neurons each, while the fourth has 128 neurons and the fifth has 32 neurons. A leaky ReLU activation function is used for all the layers. To prevent overfitting during training, a 5\% dropout rate is implemented for each hidden layer. 
The last hidden layer is fully connected to a single-neuron output, activated by a linear function, where ``1'' indicates a true event and ``0'' is a fake event. 
The DS-GAN was trained using about 1M two-pion event samples for 80K adversarial epochs, with an epoch defined as one pass through the training dataset. 
Both the generator and discriminator were trained using the Adam optimizer~\cite{kingma2014adam} with a learning rate of $10^{-5}$ and exponential decay rates for the moment estimates ($\beta1 = 0.5$, and $\beta2 = 0.9$).

\subsection{Unfolding GAN (UNF-GAN)}\label{sec:UNF-GAN}

The training process for the UNF-GAN is illustrated in Fig.~\ref{fig: GAN framework}, which depicts the variation of a typical GAN model structure consisting of a conditional generator and a discriminator.
The generator takes as input the photon energy generated by the RE-MC, along with a 100-dimensional white noise vector centered at zero with a unit standard deviation.
This combination of inputs allows the generator, implemented as a deep neural network, to transform the noise and photon energy into a minimal set of event features/variables that effectively describe the two-pion photoproduction reaction.

To strike a balance between execution time and convergence, the generator network is designed with 7 hidden dense layers. 
The number of neurons in each layer follows the sequence: 16, 32, 64, 128, 256, 512, and 1024, all of which are activated by the ReLU function.
The last hidden layer is fully connected to a 4-neuron output layer, activated by a linear function. 
This output layer represents the independent variables $M^2_{\pi^+\pi^-}$, $M^2_{p\pi^-}$, $t_{\pi^+}$, and $\alpha_{[\pi^+ p][\pi^-p']}$ that are specifically chosen to describe the reaction.

The synthetic GEN event features, generated by the conditional GAN generator, are then fed into the DS-GAN to incorporate the detector effects, and then compared to REC pseudodata obtained by passing the GEN RE-MC pseudodata through GEANT. 
The training process involved utilizing approximately 400k two-pion event samples for a duration of around 200k adversarial epochs per UNF-GAN model. 
Consistent configuration parameters for the Adam optimizer were maintained, utilizing the same settings as employed for the DS-GAN.
During the training, the generator and the discriminator engage in an adversarial competition, with both updating their parameters throughout the process. Eventually, the generator is able to generate synthetic REC samples that are indistinguishable from the REC pseudodata samples.
This means that the discriminator's ability to correctly classify whether a sample is genuine or synthetic approximates random chance.

\begin{figure}[t]
\begin{center}
\includegraphics[width=1\columnwidth]{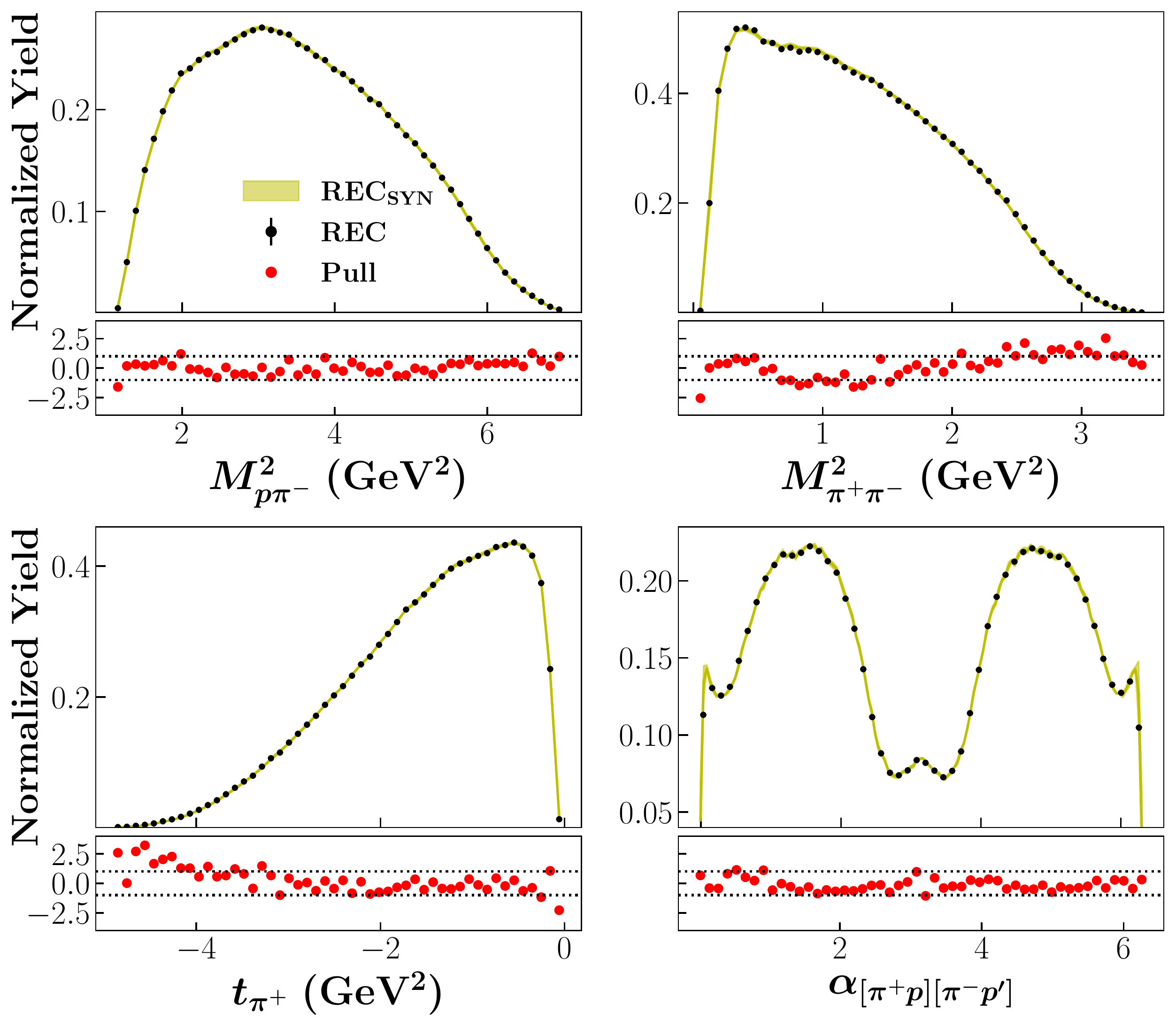}
\vspace{-0.3cm}
\caption{Training variable PS-MC REC pseudodata distributions (black points) are compared to the synthetic data produced by the DS-GAN (yellow band). The band size reflects the uncertainty estimated using the bootstrap procedure, while the pull distributions (bottom of panels) quantify the agreement between the two, with the horizontal dotted lines corresponding to $\pm 1\sigma$.}
\label{fig:DS-GAN-trainigvars}
\end{center}
\end{figure}

\begin{figure}[t]
\begin{center}
\includegraphics[width=1\columnwidth]{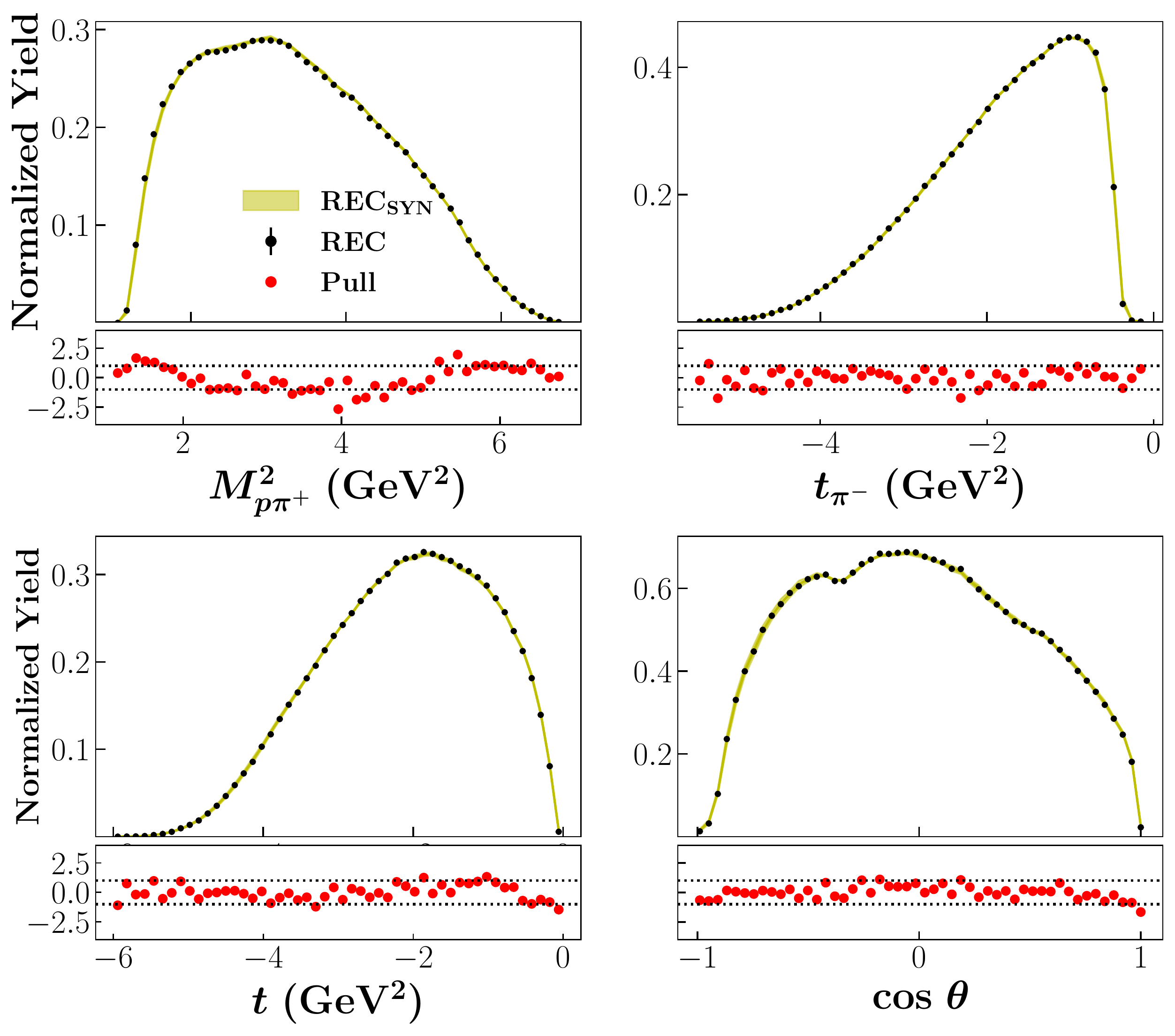}
\vspace{-0.3cm}
\caption{Same as Fig.~\ref{fig:DS-GAN-trainigvars}, but for derived quantities not used in the training process.}
\label{fig:DS-GAN-derivedvars}
\end{center}
\end{figure}

\subsection{Uncertainty quantification}
\label{sec:uq}

As neural networks become increasingly employed in physics analysis, it becomes crucial to accurately assess the reliability of ML predictions. 
The statistics of the synthetic samples can be made arbitrarily high, so that there is no need to consider a statistical uncertainty. 
However, it is important to quantify the systematic uncertainty related to the training procedure, and for this a bootstrap resampling technique was employed. 
For the DS-GAN, the procedure involved training a total of 20 neural networks independently from the beginning.
Each one was trained on a different  random sample set drawn from the original dataset with replacement, resulting in datasets of the same size but with potentially different observations.

For the UNF-GAN a similar procedure was adopted, with 20 different networks trained independently using the same bootstrap resampling technique. 
Moreover, each of the 20 UNF-GANs used a different DS-GAN of the 20 discussed above.
In this way, the systematic uncertainties associated with the DS- and UNF-GANs 
are effectively combined.
While it is possible that using a higher number of bootstraps could potentially lead to more precise uncertainty estimates, we found that training 20 GANs provided reasonably stable and consistent results.

It is important to note that the specific number of bootstraps can vary depending on the characteristics of the problem, available data, and desired level of uncertainty quantification. 
In this particular case, 20 bootstraps were deemed sufficient for accurately capturing and quantifying the uncertainties associated with the observables.
Furthermore, changing the network architecture was not essential because the convergence we achieved, along with the estimated error and uncertainty quantification, clearly indicate that this architecture is capable of accurately reproducing the data without introducing further systematic uncertainties.

\begin{figure}[t]
\begin{center}
\includegraphics[width=0.95\columnwidth]{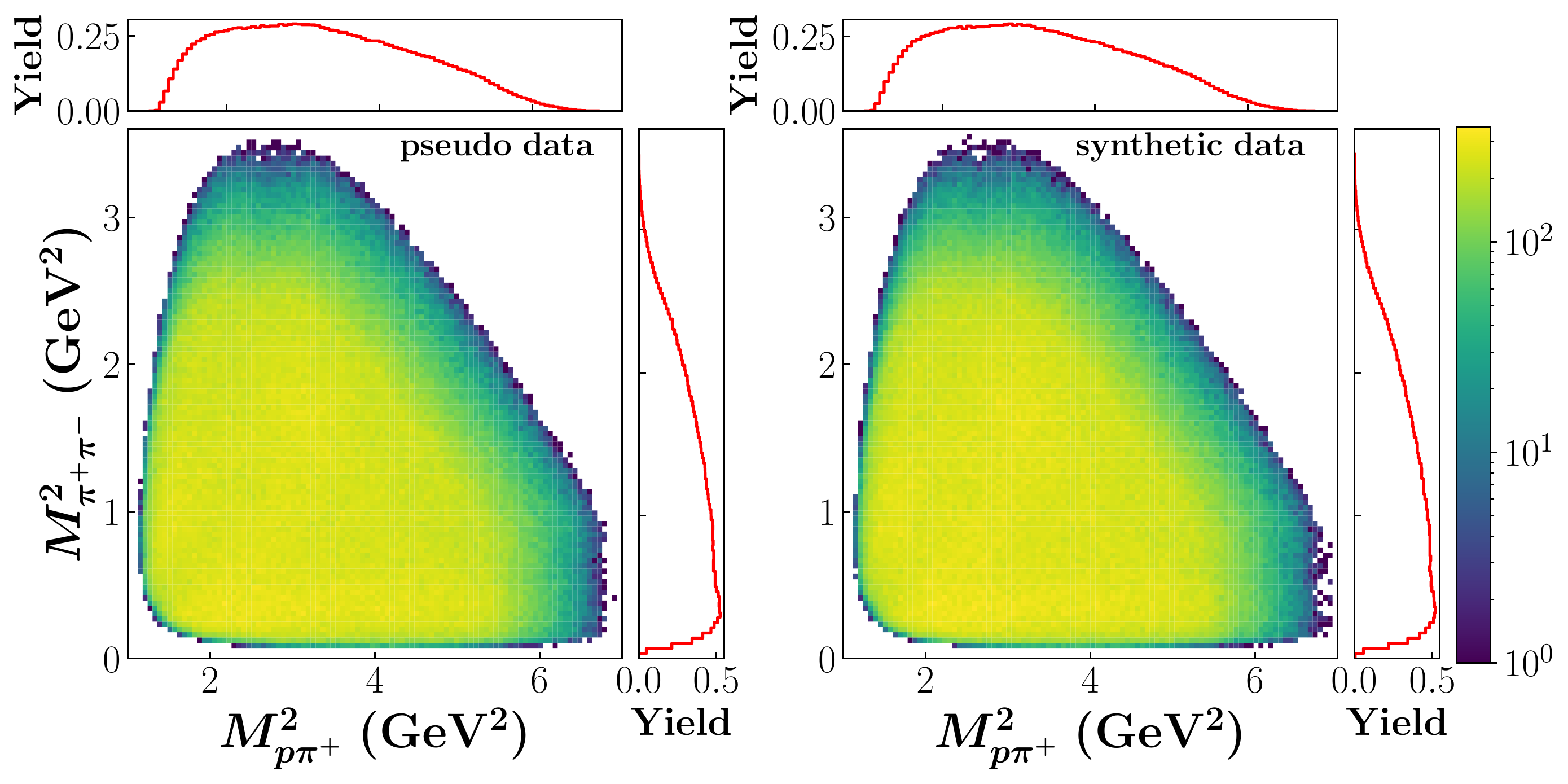}
\vspace{-0.3cm}
\caption{2D distributions of $\pi^+\pi^-$ and $p\pi^+$ invariant masses squared in PS-MC REC pseudodata (left) and synthetic data output of one of the 20 DS-GANs (right).}
\label{fig:DS-GAN-2D}
\end{center}
\end{figure}

\section{Results}
\label{sec:closure}

In this section we now discuss the DS-GAN and UNF-GAN performance, comparing synthetic to the REC and GEN pseudodata. 
We use the nomenclature REC$_\text{SYN}$ and GEN$_\text{SYN}$ to indicate synthetic data at the detector and vertex levels, respectively.
To visualize the comparison, we build marginal 1D and 2D histograms for some kinematic variables.
To show that correlations are correctly accounted for, we also study the distribution of one variable in some slices of the other variables.

Synthetic data are generated with the 
bootstrap procedure detailed in Sec.~\ref{sec:uq}, so that the standard deviation $\sigma_\text{SYN}$ corresponds to the systematic uncertainty. 
In all our results, the average $\mu_\text{SYN}$ is shown as a solid line, together with an error band of width $\pm 1\sigma_\text{SYN}$, while pseudodata are represented by dots with their statistical uncertainty $\sigma_\text{pseudodata}$.
To quantify the level of agreement between the synthetic data and pseudodata, we plot the pull for each bin, defined as
\begin{equation}
{\rm pull}\ =\ 
\frac{\mu_\text{SYN}-\mu_\text{pseudodata}}{\sqrt{\sigma^2_\text{SYN}+\sigma^2_\text{pseudodata}}},
\end{equation}
where $\mu_\text{pseudodata}$ denotes the mean of the pseudodata.
%

\subsection{DS-GAN}

The DS-GAN is trained on four independent variables: the invariant masses $M_{p\pi^-}^2$ and $M_{\pi^+\pi^-}^2$, $t_{\pi^+}$, and the angle $\alpha_{[\pi^+ p][\pi^-p']}$. 
The comparison between REC$_{\text{SYN}}$ and pseudodata PS-MC REC distributions is shown Fig.~\ref{fig:DS-GAN-trainigvars}.
In Fig.~\ref{fig:DS-GAN-derivedvars} the comparison is extended to other physics-relevant distributions not used in the training and derived from the four above-mentioned variables, namely $M_{p\pi^+}^2$, $t_{\pi^-}$, $t$, and $\cos \theta$. 
The agreement, quantified by the pull distributions shown at the bottom of each plot, is remarkable, in both cases, with most of the points lying within $1\sigma$.
This indicates that the DS-GAN is indeed able to learn the CLAS detector effects.
Bidimensional distributions from MC and synthetic data are shown in Fig.~\ref{fig:DS-GAN-2D}.

\begin{figure}[t!]
\begin{center}
\includegraphics[width=1\columnwidth]{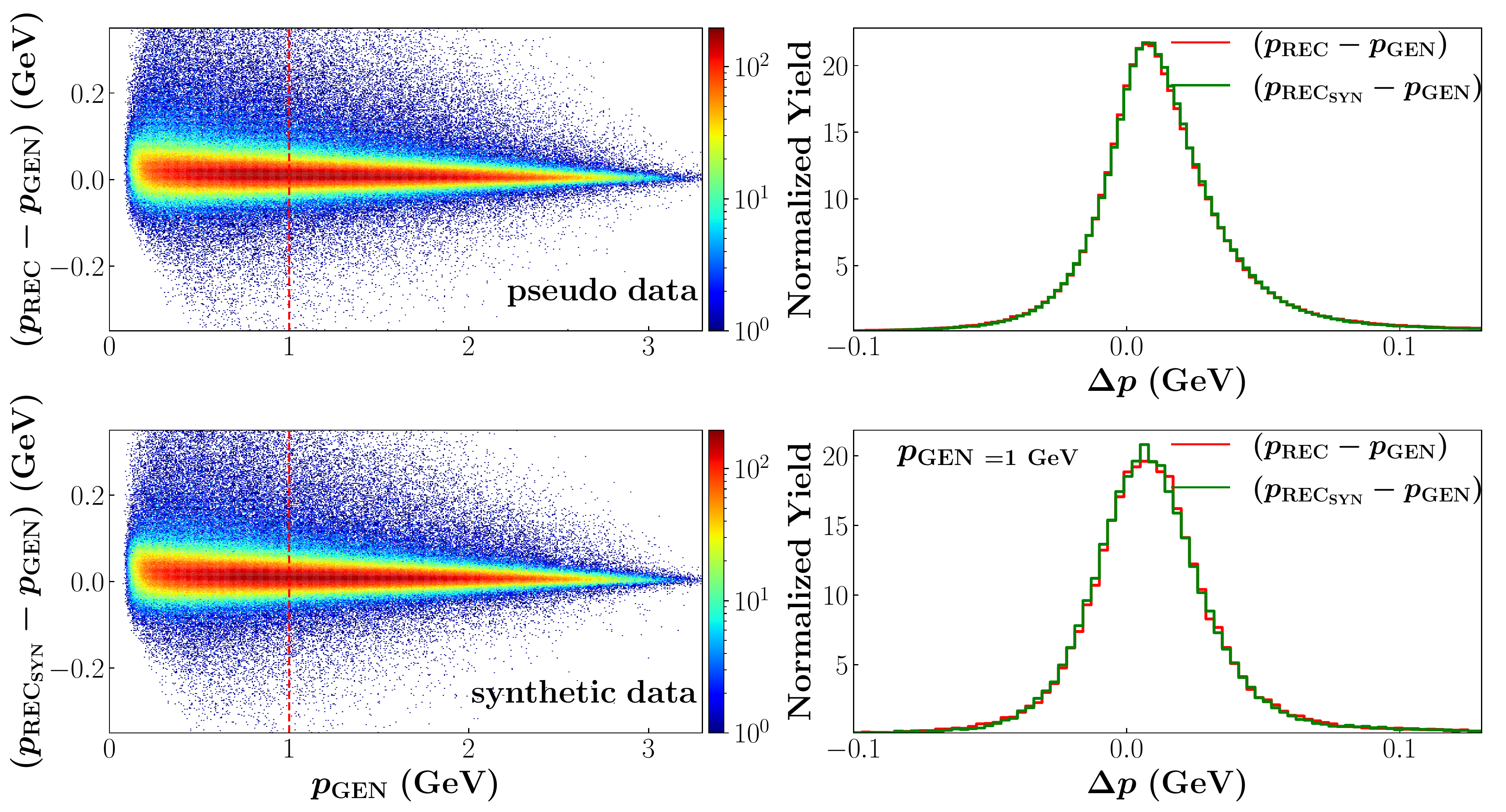}
\vspace{-0.3cm}
\caption{Left: Comparison of $\pi^+$ absolute momentum resolution for PS-MC REC pseudodata (top) and DS-GAN synthetic data (bottom) as a function of $p_\text{GEN}$. Right: The same but integrated over $p_\text{GEN}$ (upper) or for a narrow slice at $p_\text{GEN} = 1$~GeV (lower).}
\label{fig:P_rec-P_gen}
\end{center}
\end{figure}

The $\pi^+$ absolute momentum resolution as obtained from pseudodata (REC$-$GEN) is shown in Fig.~\ref{fig:P_rec-P_gen}, along with synthetic data (REC$_{\text{SYN}}-$GEN).
The two distributions are in very good agreement, indicating that synthetic data incorporate the correct resolution of the detector. 
Similar results hold for other kinematic variables of all particles. 

These comparisons demonstrate the ability of the DS-GAN to learn and reproduce detector effects in a multidimensional space, even in the tails of the distributions.
This confirms that generative models can indeed be used as an efficient and fast proxy for more computational expensive GEANT simulations~\cite{bellagente2020gan}.

\begin{figure}[hbt!]
\begin{center}
\includegraphics[width=0.95\columnwidth]{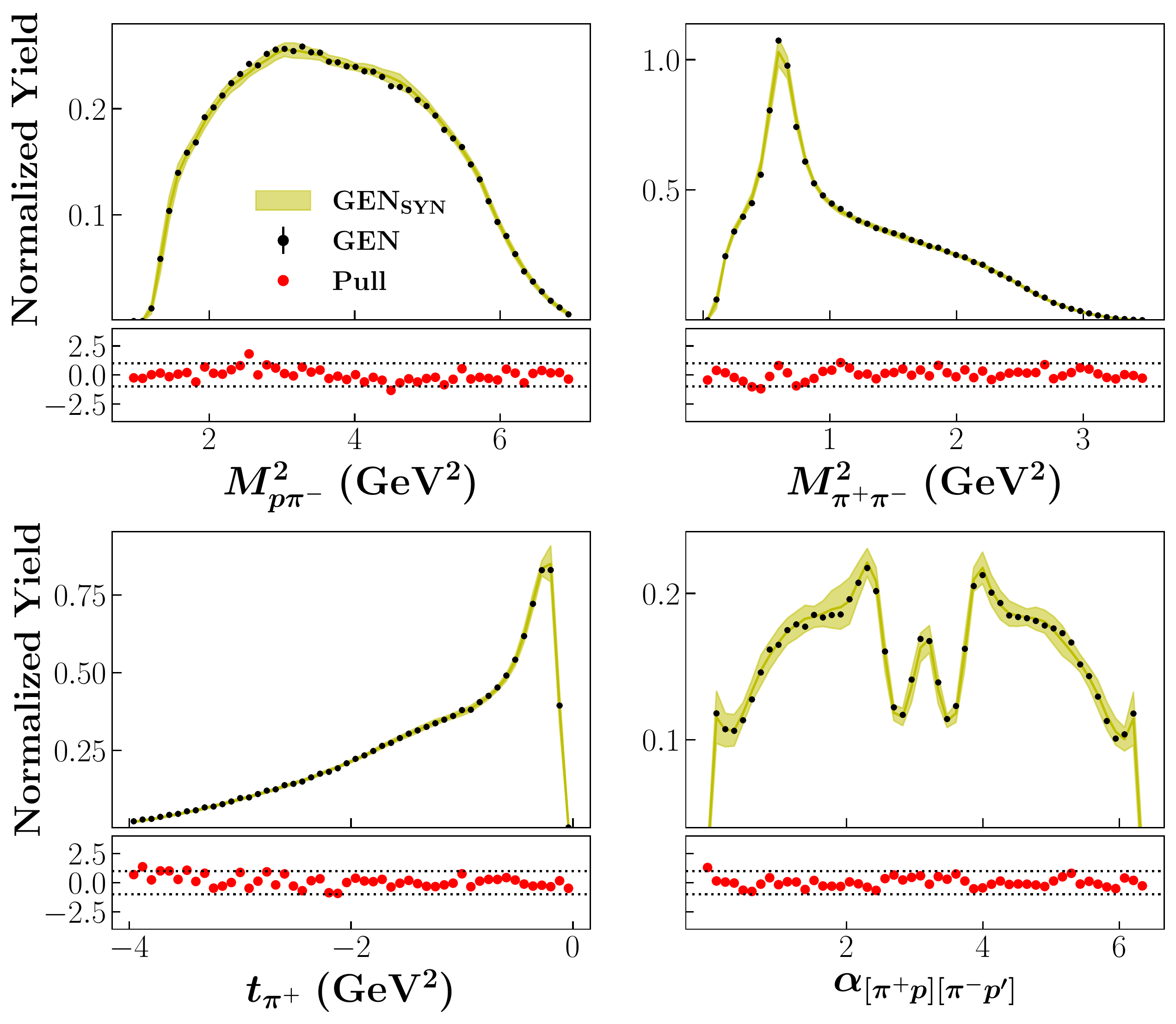}
\vspace{-0.3cm}
\caption{Comparison of GEN and GEN$_{\text{SYN}}$ distributions for the four training variables obtained by the UNF-GAN. Pseudodata were generated using RE-MC.}
\label{fig:UNF-GAN-training_single_GAN}
\end{center}
\end{figure}

\subsection{UNF-GAN}

As described in Sec~\ref{sec:UNF-GAN}, the final step in the closure test is to use REC RE-MC pseudodata to train the UNF-GAN, extract the GEN$_{\text{SYN}}$ distributions, and compare them with GEN pseudodata. 
Figure~\ref{fig:UNF-GAN-training_single_GAN} shows the comparison between GEN and GEN$_{\text{SYN}}$ for the four training variables. 
We can see a very good agreement between pseudo- and synthetic data at the vertex level, despite the fact that the UNF-GAN was trained on detector-level pseudodata. 
This clearly demonstrates the success of the unfolding procedure.

\begin{figure}[t]
\begin{center}
\includegraphics[width=0.95\columnwidth]{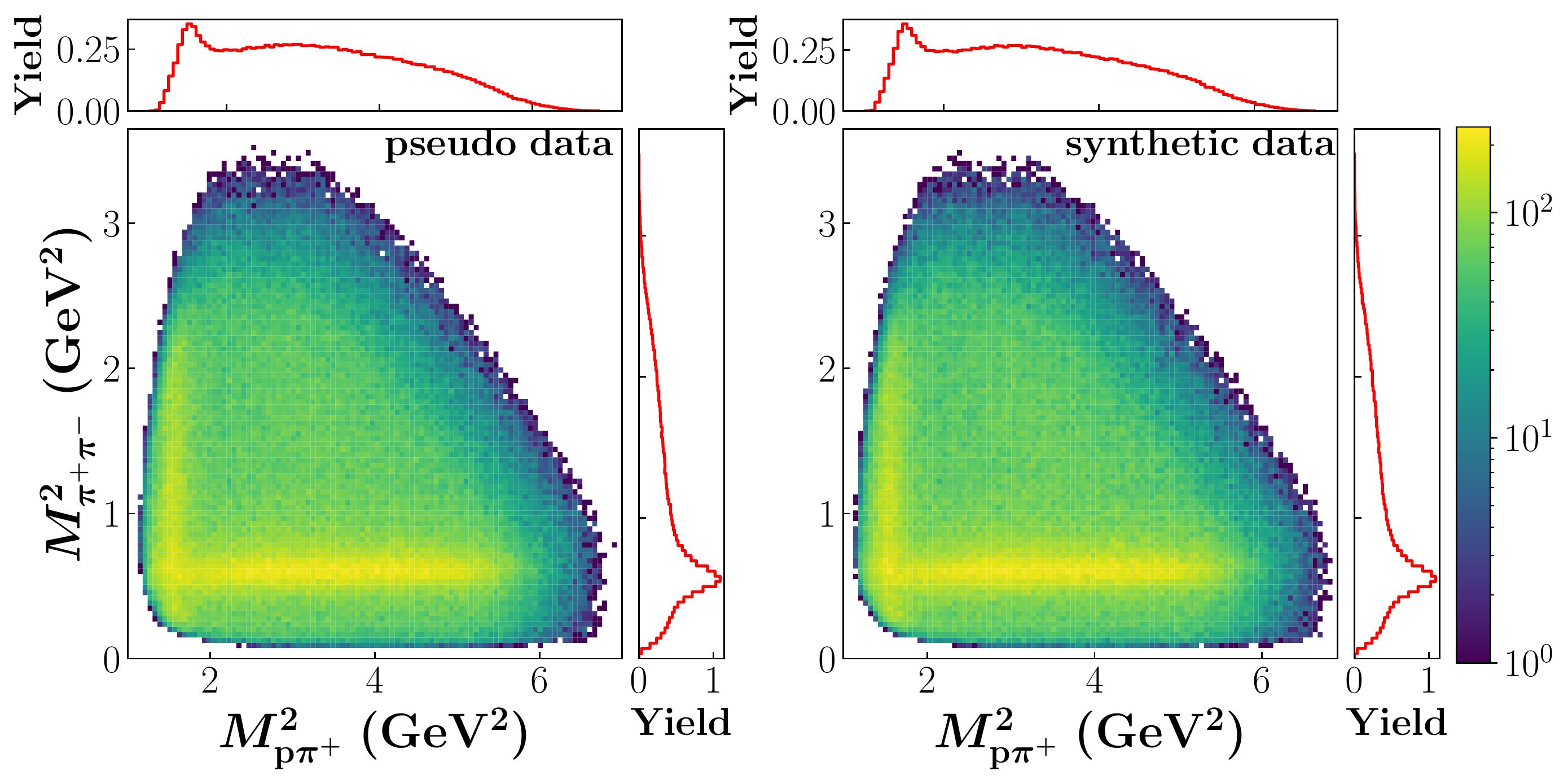}
\vspace{-0.3cm}
\caption{2D distributions of $\pi^+\pi^-$ and $p\pi^+$ invariant masses squared in GEN RE-MC pseudodata (left) and synthetic data output of one of the 20 UNF-GANs (right).} 
\label{fig:UNF-GAN-2D}
\end{center}
\end{figure}

\begin{figure}[t]
\begin{center}
\includegraphics[width=0.95\columnwidth]{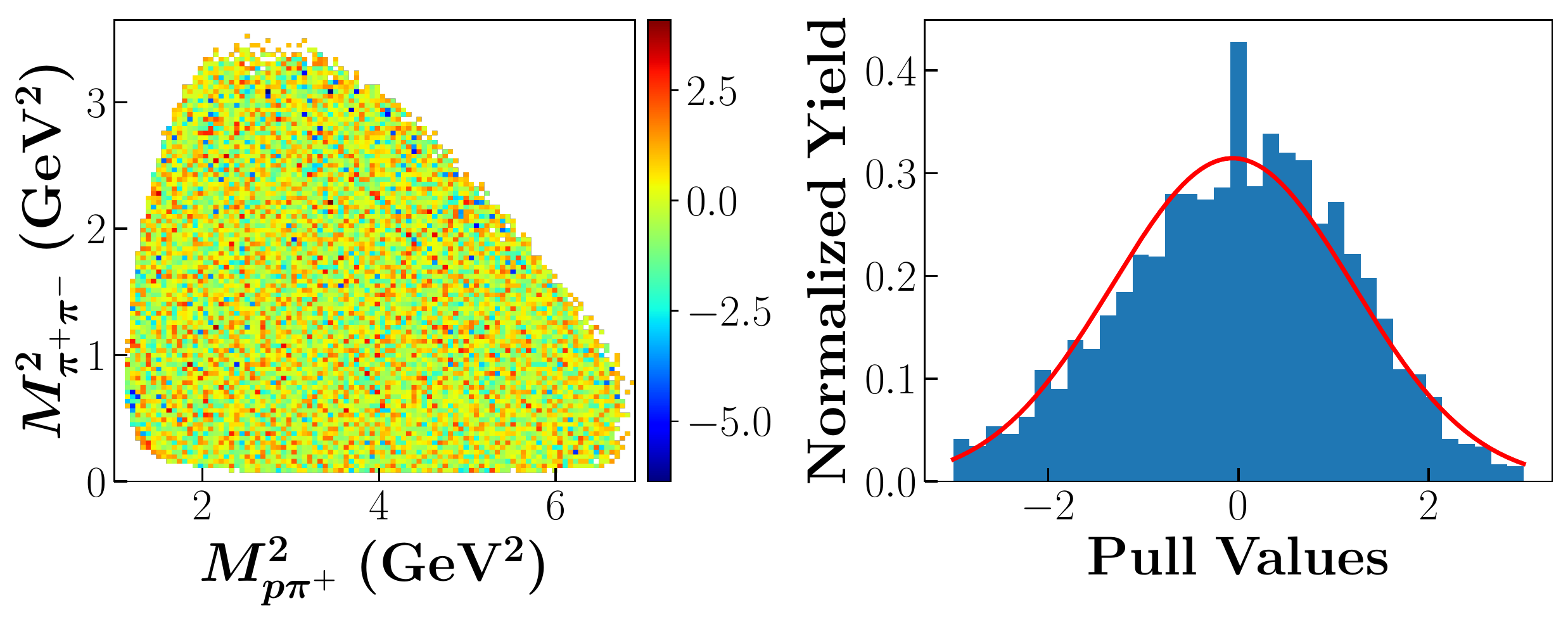}
\vspace{-0.3cm}
\caption{Left panel: 2D pulls of the distributions shown in Fig.~\ref{fig:UNF-GAN-2D}. Color palette shows the value of the pull. Right panel: histogram of the 2D pulls with a fitted Gaussian distribution centered at zero with a standard deviation of 1.3.}
\label{fig:UNF-GAN-pulls-2D}
\end{center}
\end{figure}

Moreover, the vast majority of pulls lie within $\pm 1\sigma$, indicating that the uncertainty quantification is appropriate. 
The key point of this closure test is to demonstrate that synthetic data maintain the correlations of the original pseudodata. 
We checked that this is indeed the case: in Fig.~\ref{fig:UNF-GAN-2D} we display an example of 2D distributions featuring strong correlations. 
We give a quantitative determination of the success of the procedure by calculating the pulls, shown in Fig.~\ref{fig:UNF-GAN-pulls-2D}, which turn out to be normally distributed, as expected.

\begin{figure}[t]
\begin{center}
\includegraphics[width=1\columnwidth]{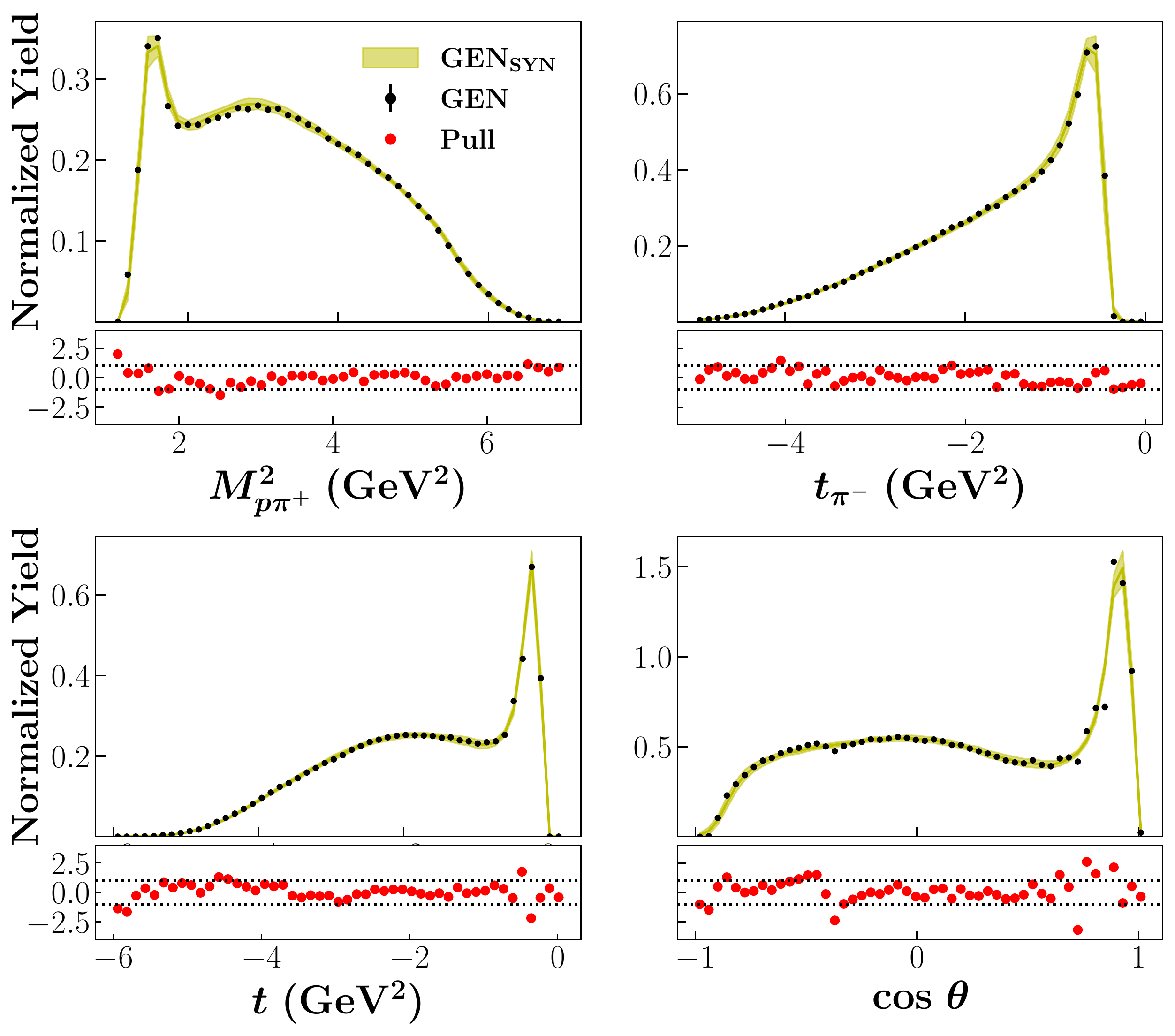}
\caption{
Comparison of GEN and GEN$_{\text{SYN}}$ for four  derived variables $M^2_{p\pi^+}$, $t_{\pi^-}$, $t$, and $\cos{\theta}$. Events were generated using RE-MC. 
}
\label{fig:UNF-GAN-derivedvarsInv}
\end{center}
\end{figure}

\begin{figure}[bt]
\begin{center}
\vspace{-0.3cm}
\includegraphics[width=0.95\columnwidth]{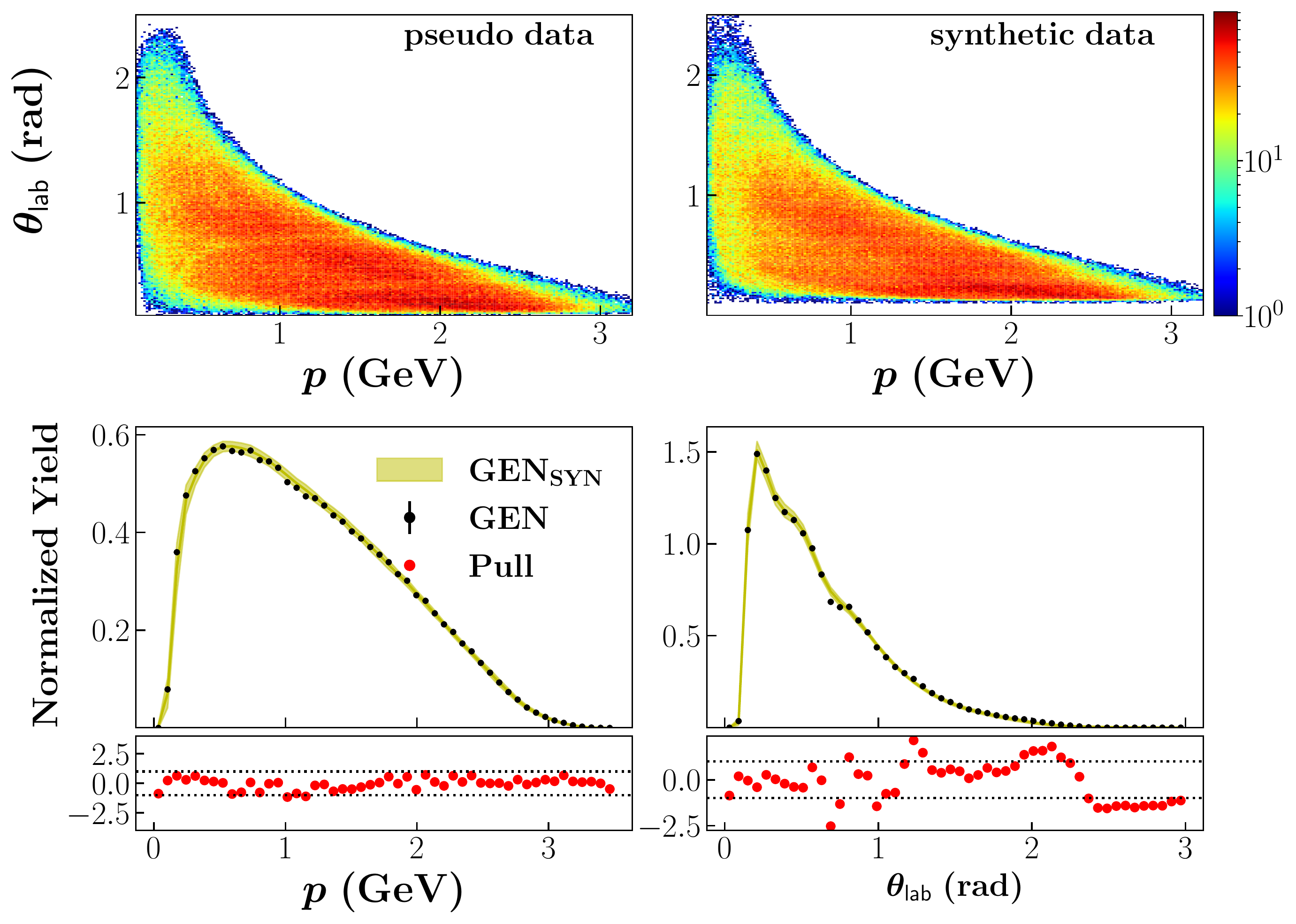}
\caption{Comparison of GEN and GEN$_\text{SYN}$ for $p$ momentum components and $\theta_\text{lab}$ in the laboratory reference frame using RE-MC data.}
\label{fig:UNF-GAN-derivedvarsLab2d}
\end{center}
\end{figure}

The good agreement and preservation of correlations remains valid for derived kinematic variables that were not used for training. 
Examples are shown in Fig.~\ref{fig:UNF-GAN-derivedvarsInv} for invariant and CM variables, and in Fig.~\ref{fig:UNF-GAN-derivedvarsLab2d} for variables in the lab frame.
It is worth noting that in the lab frame the GEN pseudodata exhibits sharp features due to detector acceptance. 
These features cannot be properly captured by the GANs, which is trained on invariant variables. Even so, this results in a $\lesssim 2\sigma$ local discrepancy in the 1D projections. 
If better agreement is needed, lab frame variables can be added to the training set.

\begin{figure}[t]
\begin{center}
\includegraphics[width=0.95\columnwidth]{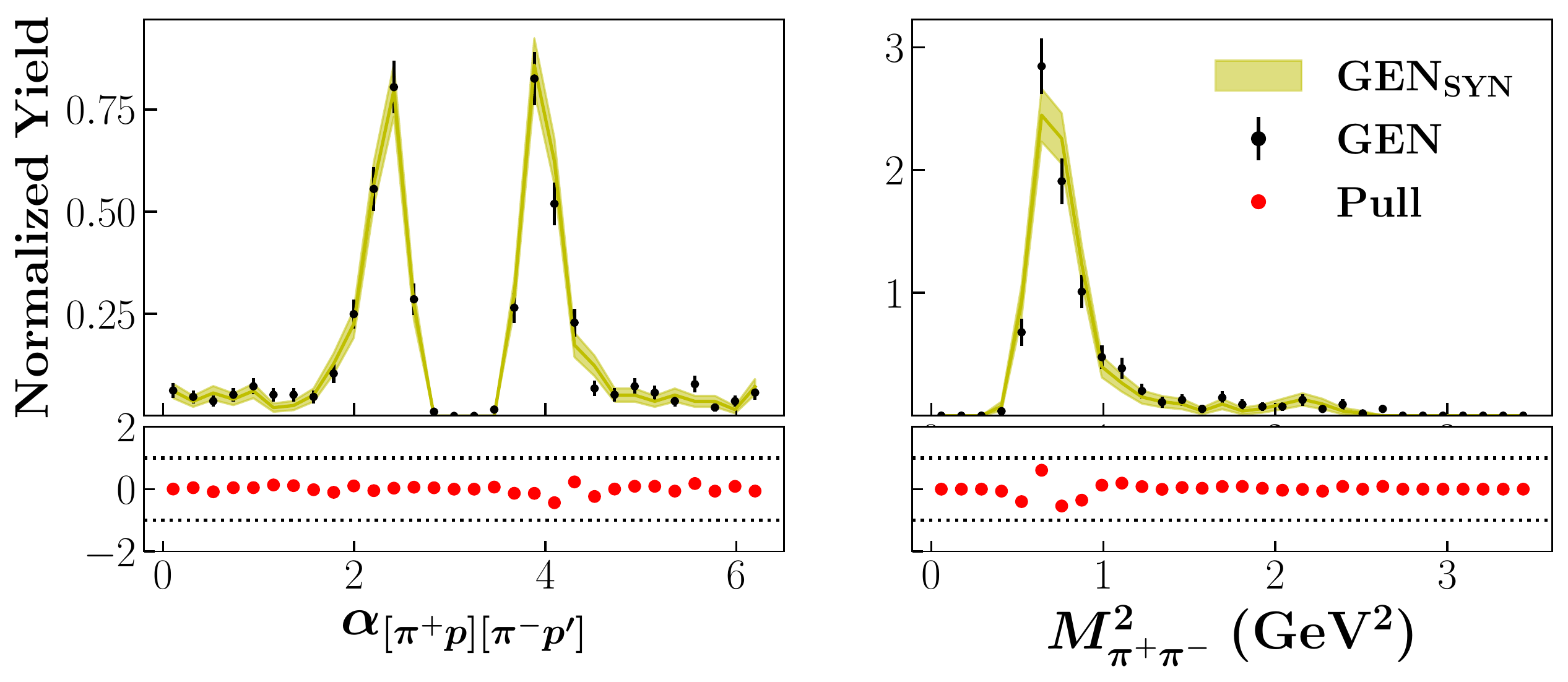}
\caption{1D histograms for fixed slice of the other variables ($2.55 < W < 2.60\text{~GeV}$,  $-0.7 < t_{\pi^+} < -0.3\text{~GeV}^2$, $2.5 < M_{p\pi^-}^2 < 3.3\text{~GeV}^2$). Left panel: $\alpha_{[\pi^+ p][\pi^-p']}$ distribution for  $0.6 < M_{\pi^+\pi^-}^2 < 0.9\text{~GeV}^2$. Right panel: $M_{\pi^+\pi^-}^2$ distribution for $2.0 < \alpha_{[\pi^+ p][\pi^-p']} < 2.5$.
}
\label{fig:UNF-GAN-slice}
\end{center}
\end{figure}

Finally, in Fig.~\ref{fig:UNF-GAN-slice} we compare 1D distributions in a given bin of the other variables. 
The success of this test shows that correlations underlying the multidifferential cross section are correctly reproduced in the synthetic datasets.

\section{Conclusions and outlook}
\label{sec:conclusions}

One of the central results of this paper is the demonstration that a generative adversarial network can be used to reproduce a realistic multibody physics reaction. 
As a case study, we have used two-pion photoproduction in the kinematics of the Jefferson Lab CLAS \geleven experiment.
This process represents an ideal test case, where several baryon and meson production mechanisms overlap, resulting in rich and complex observable distributions.
The nonuniformity of the CLAS detector response further adds complication to the challenge.

In order to validate the framework, we have performed a closure test to demonstrate that synthetic data correctly reproduce the multidifferential cross section preserving correlations between kinematic variables. 
Detector effects were also correctly unfolded by the procedure.
We deployed two MC event generators, one distributed according to pure phase space, and the other incorporating a realistic physics model.
Generated pseudodata were fed into a GEANT-based detector model to realistically take into account the detector response.
Phase-space pseudodata were used to train a GAN-based proxy to learn the detector effects, and realistic pseudodata were then used to train the unfolding GAN and generate synthetic copies of MC events.

The uncertainty quantification of the entire procedure was assessed by combining a bootstrap for the two NNs. 
Comparison between the true and GAN-generated samples demonstrated that, within the quoted systematic error, the NN is able to reproduce training and derived kinematic variables, as well as to unfold the detector effects in multiple dimensions.

This work represents a first step towards a full AI-supported analysis of CLAS exclusive two-pion photoproduction data. 
It demonstrates that the same analysis framework, trained on CLAS data, can provide a synthetic copy of the experimental data, preserving correlations between kinematic variables and unfolding the detector effects.
Physics interpretation in term of production mechanisms, separating different contributions and extracting resonance parameters from the unfolded data, will follow.
An extension of this framework to include the different topologies and extrapolating in a controlled (albeit model-dependent) outside detector acceptance is also in progress.

\acknowledgements

We thank J.~Qiu for helpful discussions.
This work was supported by the Jefferson Lab LDRD project No.~LDRD19-13 and No.~LDRD20-18, and in part by the U.S. Department of Energy contract DE-AC05-06OR23177, under which Jefferson Science Associates, LLC, manages and operates Jefferson Lab. ANHB is supported by the DFG through the Research Unit FOR 2926 (project number 409651613). TA was supported by a Ph.D. scholarship from Al-Baha University, Saudi Arabia. The work of NS was supported by the DOE, Office of Science, Office of Nuclear Physics in the Early Career Program. This work contributes to the aims of the U.S. Department of Energy ExoHad Topical Collaboration, contract DE-SC0023598.

\bibliographystyle{apsrev4-1}
\bibliography{biblio}

\begin{thebibliography}{58}%
\makeatletter
\providecommand \@ifxundefined [1]{%
 \@ifx{#1\undefined}
}%
\providecommand \@ifnum [1]{%
 \ifnum #1\expandafter \@firstoftwo
 \else \expandafter \@secondoftwo
 \fi
}%
\providecommand \@ifx [1]{%
 \ifx #1\expandafter \@firstoftwo
 \else \expandafter \@secondoftwo
 \fi
}%
\providecommand \natexlab [1]{#1}%
\providecommand \enquote  [1]{``#1''}%
\providecommand \bibnamefont  [1]{#1}%
\providecommand \bibfnamefont [1]{#1}%
\providecommand \citenamefont [1]{#1}%
\providecommand \href@noop [0]{\@secondoftwo}%
\providecommand \href [0]{\begingroup \@sanitize@url \@href}%
\providecommand \@href[1]{\@@startlink{#1}\@@href}%
\providecommand \@@href[1]{\endgroup#1\@@endlink}%
\providecommand \@sanitize@url [0]{\catcode `\\12\catcode `\$12\catcode
  `\&12\catcode `\#12\catcode `\^12\catcode `\_12\catcode `\%12\relax}%
\providecommand \@@startlink[1]{}%
\providecommand \@@endlink[0]{}%
\providecommand \url  [0]{\begingroup\@sanitize@url \@url }%
\providecommand \@url [1]{\endgroup\@href {#1}{\urlprefix }}%
\providecommand \urlprefix  [0]{URL }%
\providecommand \Eprint [0]{\href }%
\providecommand \doibase [0]{http://dx.doi.org/}%
\providecommand \selectlanguage [0]{\@gobble}%
\providecommand \bibinfo  [0]{\@secondoftwo}%
\providecommand \bibfield  [0]{\@secondoftwo}%
\providecommand \translation [1]{[#1]}%
\providecommand \BibitemOpen [0]{}%
\providecommand \bibitemStop [0]{}%
\providecommand \bibitemNoStop [0]{.\EOS\space}%
\providecommand \EOS [0]{\spacefactor3000\relax}%
\providecommand \BibitemShut  [1]{\csname bibitem#1\endcsname}%
\let\auto@bib@innerbib\@empty
\bibitem [{\citenamefont {Battaglieri}\ \emph {et~al.}(2004)\citenamefont
  {Battaglieri}, \citenamefont {De~Vita}, \citenamefont {Kubarovsky},
  \citenamefont {Price}, \citenamefont {Stoler},\ and\ \citenamefont
  {Weygand}}]{g11}%
  \BibitemOpen
  \bibfield  {author} {\bibinfo {author} {\bibfnamefont {M.}~\bibnamefont
  {Battaglieri}}, \bibinfo {author} {\bibfnamefont {R.}~\bibnamefont
  {De~Vita}}, \bibinfo {author} {\bibfnamefont {V.}~\bibnamefont {Kubarovsky}},
  \bibinfo {author} {\bibfnamefont {J.}~\bibnamefont {Price}}, \bibinfo
  {author} {\bibfnamefont {P.}~\bibnamefont {Stoler}}, \ and\ \bibinfo {author}
  {\bibfnamefont {D.}~\bibnamefont {Weygand}},\ }\href
  {http://www.jlab.org/exp_prog/proposals/04/PR04-017.pdf} {\enquote {\bibinfo
  {title} {Study of pentaquark states in photoproduction off protons},}\ }
  (\bibinfo {year} {2004})\BibitemShut {NoStop}%
\bibitem [{\citenamefont {Alanazi}\ \emph {et~al.}(2022)\citenamefont
  {Alanazi}, \citenamefont {Ambrozewicz}, \citenamefont {Battaglieri},
  \citenamefont {Hiller~Blin}, \citenamefont {Kuchera}, \citenamefont {Li},
  \citenamefont {Liu}, \citenamefont {McClellan}, \citenamefont {Melnitchouk},
  \citenamefont {Pritchard}, \citenamefont {Robertson}, \citenamefont {Sato},
  \citenamefont {Strauss},\ and\ \citenamefont {Velasco}}]{Alanazi:2020jod}%
  \BibitemOpen
  \bibfield  {author} {\bibinfo {author} {\bibfnamefont {Y.}~\bibnamefont
  {Alanazi}}, \bibinfo {author} {\bibfnamefont {P.}~\bibnamefont
  {Ambrozewicz}}, \bibinfo {author} {\bibfnamefont {M.}~\bibnamefont
  {Battaglieri}}, \bibinfo {author} {\bibfnamefont {A.~N.}\ \bibnamefont
  {Hiller~Blin}}, \bibinfo {author} {\bibfnamefont {M.~P.}\ \bibnamefont
  {Kuchera}}, \bibinfo {author} {\bibfnamefont {Y.}~\bibnamefont {Li}},
  \bibinfo {author} {\bibfnamefont {T.}~\bibnamefont {Liu}}, \bibinfo {author}
  {\bibfnamefont {R.~E.}\ \bibnamefont {McClellan}}, \bibinfo {author}
  {\bibfnamefont {W.}~\bibnamefont {Melnitchouk}}, \bibinfo {author}
  {\bibfnamefont {E.}~\bibnamefont {Pritchard}}, \bibinfo {author}
  {\bibfnamefont {M.}~\bibnamefont {Robertson}}, \bibinfo {author}
  {\bibfnamefont {N.}~\bibnamefont {Sato}}, \bibinfo {author} {\bibfnamefont
  {R.}~\bibnamefont {Strauss}}, \ and\ \bibinfo {author} {\bibfnamefont
  {L.}~\bibnamefont {Velasco}},\ }\href {\doibase 10.1103/PhysRevD.106.096002}
  {\bibfield  {journal} {\bibinfo  {journal} {Phys. Rev. D}\ }\textbf {\bibinfo
  {volume} {106}},\ \bibinfo {pages} {096002} (\bibinfo {year} {2022})},\
  \Eprint {http://arxiv.org/abs/2008.03151} {arXiv:2008.03151 [hep-ph]}
  \BibitemShut {NoStop}%
\bibitem [{\citenamefont {Brun}\ \emph {et~al.}(1987)\citenamefont {Brun},
  \citenamefont {Bruyant}, \citenamefont {Maire}, \citenamefont {McPherson},\
  and\ \citenamefont {Zanarini}}]{Brun:1987ma}%
  \BibitemOpen
  \bibfield  {author} {\bibinfo {author} {\bibfnamefont {R.}~\bibnamefont
  {Brun}}, \bibinfo {author} {\bibfnamefont {F.}~\bibnamefont {Bruyant}},
  \bibinfo {author} {\bibfnamefont {M.}~\bibnamefont {Maire}}, \bibinfo
  {author} {\bibfnamefont {A.~C.}\ \bibnamefont {McPherson}}, \ and\ \bibinfo
  {author} {\bibfnamefont {P.}~\bibnamefont {Zanarini}},\ }\href
  {https://cds.cern.ch/record/1119728} {\emph {\bibinfo {title} {{GEANT 3:
  user's guide Geant 3.10, Geant 3.11; rev. version}}}}\ (\bibinfo  {publisher}
  {CERN},\ \bibinfo {address} {Geneva},\ \bibinfo {year} {1987})\BibitemShut
  {NoStop}%
\bibitem [{\citenamefont {Capstick}\ and\ \citenamefont
  {Roberts}(2000)}]{Capstick:2000qj}%
  \BibitemOpen
  \bibfield  {author} {\bibinfo {author} {\bibfnamefont {S.}~\bibnamefont
  {Capstick}}\ and\ \bibinfo {author} {\bibfnamefont {W.}~\bibnamefont
  {Roberts}},\ }\href {\doibase 10.1016/S0146-6410(00)00109-5} {\bibfield
  {journal} {\bibinfo  {journal} {Prog. Part. Nucl. Phys.}\ }\textbf {\bibinfo
  {volume} {45}},\ \bibinfo {pages} {S241} (\bibinfo {year} {2000})},\ \Eprint
  {http://arxiv.org/abs/nucl-th/0008028} {arXiv:nucl-th/0008028} \BibitemShut
  {NoStop}%
\bibitem [{\citenamefont {Klempt}\ and\ \citenamefont
  {Richard}(2010)}]{Klempt:2009pi}%
  \BibitemOpen
  \bibfield  {author} {\bibinfo {author} {\bibfnamefont {E.}~\bibnamefont
  {Klempt}}\ and\ \bibinfo {author} {\bibfnamefont {J.-M.}\ \bibnamefont
  {Richard}},\ }\href {\doibase 10.1103/RevModPhys.82.1095} {\bibfield
  {journal} {\bibinfo  {journal} {Rev. Mod. Phys.}\ }\textbf {\bibinfo {volume}
  {82}},\ \bibinfo {pages} {1095} (\bibinfo {year} {2010})},\ \Eprint
  {http://arxiv.org/abs/0901.2055} {arXiv:0901.2055 [hep-ph]} \BibitemShut
  {NoStop}%
\bibitem [{\citenamefont {Giannini}\ and\ \citenamefont
  {Santopinto}(2015)}]{Giannini:2015zia}%
  \BibitemOpen
  \bibfield  {author} {\bibinfo {author} {\bibfnamefont {M.~M.}\ \bibnamefont
  {Giannini}}\ and\ \bibinfo {author} {\bibfnamefont {E.}~\bibnamefont
  {Santopinto}},\ }\href {\doibase 10.6122/CJP.20150120} {\bibfield  {journal}
  {\bibinfo  {journal} {Chin. J. Phys.}\ }\textbf {\bibinfo {volume} {53}},\
  \bibinfo {pages} {020301} (\bibinfo {year} {2015})},\ \Eprint
  {http://arxiv.org/abs/1501.03722} {arXiv:1501.03722 [nucl-th]} \BibitemShut
  {NoStop}%
\bibitem [{\citenamefont {Qin}\ \emph {et~al.}(2019)\citenamefont {Qin},
  \citenamefont {Roberts},\ and\ \citenamefont {Schmidt}}]{Qin:2019hgk}%
  \BibitemOpen
  \bibfield  {author} {\bibinfo {author} {\bibfnamefont {S.-X.}\ \bibnamefont
  {Qin}}, \bibinfo {author} {\bibfnamefont {C.~D.}\ \bibnamefont {Roberts}}, \
  and\ \bibinfo {author} {\bibfnamefont {S.~M.}\ \bibnamefont {Schmidt}},\
  }\href {\doibase 10.1007/s00601-019-1488-x} {\bibfield  {journal} {\bibinfo
  {journal} {Few Body Syst.}\ }\textbf {\bibinfo {volume} {60}},\ \bibinfo
  {pages} {26} (\bibinfo {year} {2019})},\ \Eprint
  {http://arxiv.org/abs/1902.00026} {arXiv:1902.00026 [nucl-th]} \BibitemShut
  {NoStop}%
\bibitem [{\citenamefont {Chen}\ \emph {et~al.}(2019)\citenamefont {Chen},
  \citenamefont {Krein}, \citenamefont {Roberts}, \citenamefont {Schmidt},\
  and\ \citenamefont {Segovia}}]{Chen:2019fzn}%
  \BibitemOpen
  \bibfield  {author} {\bibinfo {author} {\bibfnamefont {C.}~\bibnamefont
  {Chen}}, \bibinfo {author} {\bibfnamefont {G.~I.}\ \bibnamefont {Krein}},
  \bibinfo {author} {\bibfnamefont {C.~D.}\ \bibnamefont {Roberts}}, \bibinfo
  {author} {\bibfnamefont {S.~M.}\ \bibnamefont {Schmidt}}, \ and\ \bibinfo
  {author} {\bibfnamefont {J.}~\bibnamefont {Segovia}},\ }\href {\doibase
  10.1103/PhysRevD.100.054009} {\bibfield  {journal} {\bibinfo  {journal}
  {Phys. Rev. D}\ }\textbf {\bibinfo {volume} {100}},\ \bibinfo {pages}
  {054009} (\bibinfo {year} {2019})},\ \Eprint
  {http://arxiv.org/abs/1901.04305} {arXiv:1901.04305 [nucl-th]} \BibitemShut
  {NoStop}%
\bibitem [{\citenamefont {Edwards}\ \emph {et~al.}(2011)\citenamefont
  {Edwards}, \citenamefont {Dudek}, \citenamefont {Richards},\ and\
  \citenamefont {Wallace}}]{Edwards:2011jj}%
  \BibitemOpen
  \bibfield  {author} {\bibinfo {author} {\bibfnamefont {R.~G.}\ \bibnamefont
  {Edwards}}, \bibinfo {author} {\bibfnamefont {J.~J.}\ \bibnamefont {Dudek}},
  \bibinfo {author} {\bibfnamefont {D.~G.}\ \bibnamefont {Richards}}, \ and\
  \bibinfo {author} {\bibfnamefont {S.~J.}\ \bibnamefont {Wallace}},\ }\href
  {\doibase 10.1103/PhysRevD.84.074508} {\bibfield  {journal} {\bibinfo
  {journal} {Phys. Rev. D}\ }\textbf {\bibinfo {volume} {84}},\ \bibinfo
  {pages} {074508} (\bibinfo {year} {2011})},\ \Eprint
  {http://arxiv.org/abs/1104.5152} {arXiv:1104.5152 [hep-ph]} \BibitemShut
  {NoStop}%
\bibitem [{\citenamefont {Crede}\ and\ \citenamefont
  {Roberts}(2013)}]{Crede:2013kia}%
  \BibitemOpen
  \bibfield  {author} {\bibinfo {author} {\bibfnamefont {V.}~\bibnamefont
  {Crede}}\ and\ \bibinfo {author} {\bibfnamefont {W.}~\bibnamefont
  {Roberts}},\ }\href {\doibase 10.1088/0034-4885/76/7/076301} {\bibfield
  {journal} {\bibinfo  {journal} {Rept. Prog. Phys.}\ }\textbf {\bibinfo
  {volume} {76}},\ \bibinfo {pages} {076301} (\bibinfo {year} {2013})},\
  \Eprint {http://arxiv.org/abs/1302.7299} {arXiv:1302.7299 [nucl-ex]}
  \BibitemShut {NoStop}%
\bibitem [{\citenamefont {Ireland}\ \emph {et~al.}(2020)\citenamefont
  {Ireland}, \citenamefont {Pasyuk},\ and\ \citenamefont
  {Strakovsky}}]{Ireland:2019uwn}%
  \BibitemOpen
  \bibfield  {author} {\bibinfo {author} {\bibfnamefont {D.~G.}\ \bibnamefont
  {Ireland}}, \bibinfo {author} {\bibfnamefont {E.}~\bibnamefont {Pasyuk}}, \
  and\ \bibinfo {author} {\bibfnamefont {I.}~\bibnamefont {Strakovsky}},\
  }\href {\doibase 10.1016/j.ppnp.2019.103752} {\bibfield  {journal} {\bibinfo
  {journal} {Prog. Part. Nucl. Phys.}\ }\textbf {\bibinfo {volume} {111}},\
  \bibinfo {pages} {103752} (\bibinfo {year} {2020})},\ \Eprint
  {http://arxiv.org/abs/1906.04228} {arXiv:1906.04228 [nucl-ex]} \BibitemShut
  {NoStop}%
\bibitem [{\citenamefont {Beck}\ and\ \citenamefont
  {Thoma}(2017)}]{Beck:2017wkb}%
  \BibitemOpen
  \bibfield  {author} {\bibinfo {author} {\bibfnamefont {R.}~\bibnamefont
  {Beck}}\ and\ \bibinfo {author} {\bibfnamefont {U.}~\bibnamefont {Thoma}},\
  }\href {\doibase 10.1051/epjconf/201713402001} {\bibfield  {journal}
  {\bibinfo  {journal} {EPJ Web Conf.}\ }\textbf {\bibinfo {volume} {134}},\
  \bibinfo {pages} {02001} (\bibinfo {year} {2017})}\BibitemShut {NoStop}%
\bibitem [{\citenamefont {Burkert}\ \emph {et~al.}(2022)\citenamefont
  {Burkert}, \citenamefont {Klempt},\ and\ \citenamefont
  {Thoma}}]{Burkert:2022adb}%
  \BibitemOpen
  \bibfield  {author} {\bibinfo {author} {\bibfnamefont {V.}~\bibnamefont
  {Burkert}}, \bibinfo {author} {\bibfnamefont {E.}~\bibnamefont {Klempt}}, \
  and\ \bibinfo {author} {\bibfnamefont {U.}~\bibnamefont {Thoma}},\
  }\href@noop {} {\enquote {\bibinfo {title} {{Light-quark baryons}},}\ }
  (\bibinfo {year} {2022}),\ \Eprint {http://arxiv.org/abs/2211.12906}
  {arXiv:2211.12906 [hep-ph]} \BibitemShut {NoStop}%
\bibitem [{\citenamefont {Mokeev}\ and\ \citenamefont
  {Carman}(2022)}]{Mokeev:2022xfo}%
  \BibitemOpen
  \bibfield  {author} {\bibinfo {author} {\bibfnamefont {V.~I.}\ \bibnamefont
  {Mokeev}}\ and\ \bibinfo {author} {\bibfnamefont {D.~S.}\ \bibnamefont
  {Carman}} (\bibinfo {collaboration} {CLAS}),\ }\href {\doibase
  10.1007/s00601-022-01760-2} {\bibfield  {journal} {\bibinfo  {journal} {Few
  Body Syst.}\ }\textbf {\bibinfo {volume} {63}},\ \bibinfo {pages} {59}
  (\bibinfo {year} {2022})},\ \Eprint {http://arxiv.org/abs/2202.04180}
  {arXiv:2202.04180 [nucl-ex]} \BibitemShut {NoStop}%
\bibitem [{\citenamefont {Ripani}\ \emph {et~al.}(2003)\citenamefont {Ripani}
  \emph {et~al.}}]{CLAS:2002xbv}%
  \BibitemOpen
  \bibfield  {author} {\bibinfo {author} {\bibfnamefont {M.}~\bibnamefont
  {Ripani}} \emph {et~al.} (\bibinfo {collaboration} {CLAS}),\ }\href {\doibase
  10.1103/PhysRevLett.91.022002} {\bibfield  {journal} {\bibinfo  {journal}
  {Phys. Rev. Lett.}\ }\textbf {\bibinfo {volume} {91}},\ \bibinfo {pages}
  {022002} (\bibinfo {year} {2003})},\ \Eprint
  {http://arxiv.org/abs/hep-ex/0210054} {arXiv:hep-ex/0210054} \BibitemShut
  {NoStop}%
\bibitem [{\citenamefont {Golovatch}\ \emph {et~al.}(2019)\citenamefont
  {Golovatch} \emph {et~al.}}]{CLAS:2018drk}%
  \BibitemOpen
  \bibfield  {author} {\bibinfo {author} {\bibfnamefont {E.}~\bibnamefont
  {Golovatch}} \emph {et~al.} (\bibinfo {collaboration} {CLAS}),\ }\href
  {\doibase 10.1016/j.physletb.2018.10.013} {\bibfield  {journal} {\bibinfo
  {journal} {Phys. Lett. B}\ }\textbf {\bibinfo {volume} {788}},\ \bibinfo
  {pages} {371} (\bibinfo {year} {2019})},\ \Eprint
  {http://arxiv.org/abs/1806.01767} {arXiv:1806.01767 [nucl-ex]} \BibitemShut
  {NoStop}%
\bibitem [{\citenamefont {Mokeev}\ \emph {et~al.}(2020)\citenamefont {Mokeev}
  \emph {et~al.}}]{Mokeev:2020hhu}%
  \BibitemOpen
  \bibfield  {author} {\bibinfo {author} {\bibfnamefont {V.~I.}\ \bibnamefont
  {Mokeev}} \emph {et~al.},\ }\href {\doibase 10.1016/j.physletb.2020.135457}
  {\bibfield  {journal} {\bibinfo  {journal} {Phys. Lett. B}\ }\textbf
  {\bibinfo {volume} {805}},\ \bibinfo {pages} {135457} (\bibinfo {year}
  {2020})},\ \Eprint {http://arxiv.org/abs/2004.13531} {arXiv:2004.13531
  [nucl-ex]} \BibitemShut {NoStop}%
\bibitem [{\citenamefont {Nys}\ \emph {et~al.}(2017)\citenamefont {Nys},
  \citenamefont {Mathieu}, \citenamefont {Fern\'andez-Ram\'\i{}rez},
  \citenamefont {Hiller~Blin}, \citenamefont {Jackura}, \citenamefont
  {Mikhasenko}, \citenamefont {Pilloni}, \citenamefont {Szczepaniak},
  \citenamefont {Fox},\ and\ \citenamefont {Ryckebusch}}]{JPAC:2016lnm}%
  \BibitemOpen
  \bibfield  {author} {\bibinfo {author} {\bibfnamefont {J.}~\bibnamefont
  {Nys}}, \bibinfo {author} {\bibfnamefont {V.}~\bibnamefont {Mathieu}},
  \bibinfo {author} {\bibfnamefont {C.}~\bibnamefont
  {Fern\'andez-Ram\'\i{}rez}}, \bibinfo {author} {\bibfnamefont {A.~N.}\
  \bibnamefont {Hiller~Blin}}, \bibinfo {author} {\bibfnamefont
  {A.}~\bibnamefont {Jackura}}, \bibinfo {author} {\bibfnamefont
  {M.}~\bibnamefont {Mikhasenko}}, \bibinfo {author} {\bibfnamefont
  {A.}~\bibnamefont {Pilloni}}, \bibinfo {author} {\bibfnamefont {A.~P.}\
  \bibnamefont {Szczepaniak}}, \bibinfo {author} {\bibfnamefont
  {G.}~\bibnamefont {Fox}}, \ and\ \bibinfo {author} {\bibfnamefont
  {J.}~\bibnamefont {Ryckebusch}} (\bibinfo {collaboration} {JPAC}),\ }\href
  {\doibase 10.1103/PhysRevD.95.034014} {\bibfield  {journal} {\bibinfo
  {journal} {Phys. Rev. D}\ }\textbf {\bibinfo {volume} {95}},\ \bibinfo
  {pages} {034014} (\bibinfo {year} {2017})},\ \Eprint
  {http://arxiv.org/abs/1611.04658} {arXiv:1611.04658 [hep-ph]} \BibitemShut
  {NoStop}%
\bibitem [{\citenamefont {Mathieu}\ \emph
  {et~al.}(2018{\natexlab{a}})\citenamefont {Mathieu}, \citenamefont {Nys},
  \citenamefont {Fern\'andez-Ram\'\i{}rez}, \citenamefont {Hiller~Blin},
  \citenamefont {Jackura}, \citenamefont {Pilloni}, \citenamefont
  {Szczepaniak},\ and\ \citenamefont {Fox}}]{Mathieu:2018mjw}%
  \BibitemOpen
  \bibfield  {author} {\bibinfo {author} {\bibfnamefont {V.}~\bibnamefont
  {Mathieu}}, \bibinfo {author} {\bibfnamefont {J.}~\bibnamefont {Nys}},
  \bibinfo {author} {\bibfnamefont {C.}~\bibnamefont
  {Fern\'andez-Ram\'\i{}rez}}, \bibinfo {author} {\bibfnamefont {A.~N.}\
  \bibnamefont {Hiller~Blin}}, \bibinfo {author} {\bibfnamefont
  {A.}~\bibnamefont {Jackura}}, \bibinfo {author} {\bibfnamefont
  {A.}~\bibnamefont {Pilloni}}, \bibinfo {author} {\bibfnamefont {A.~P.}\
  \bibnamefont {Szczepaniak}}, \ and\ \bibinfo {author} {\bibfnamefont
  {G.}~\bibnamefont {Fox}} (\bibinfo {collaboration} {JPAC}),\ }\href {\doibase
  10.1103/PhysRevD.98.014041} {\bibfield  {journal} {\bibinfo  {journal} {Phys.
  Rev. D}\ }\textbf {\bibinfo {volume} {98}},\ \bibinfo {pages} {014041}
  (\bibinfo {year} {2018}{\natexlab{a}})},\ \Eprint
  {http://arxiv.org/abs/1806.08414} {arXiv:1806.08414 [hep-ph]} \BibitemShut
  {NoStop}%
\bibitem [{\citenamefont {Mathieu}\ \emph {et~al.}(2020)\citenamefont
  {Mathieu}, \citenamefont {Pilloni}, \citenamefont {Albaladejo}, \citenamefont
  {Bibrzycki}, \citenamefont {Celentano}, \citenamefont
  {Fern\'andez-Ram\'\i{}rez},\ and\ \citenamefont
  {Szczepaniak}}]{Mathieu:2020zpm}%
  \BibitemOpen
  \bibfield  {author} {\bibinfo {author} {\bibfnamefont {V.}~\bibnamefont
  {Mathieu}}, \bibinfo {author} {\bibfnamefont {A.}~\bibnamefont {Pilloni}},
  \bibinfo {author} {\bibfnamefont {M.}~\bibnamefont {Albaladejo}}, \bibinfo
  {author} {\bibfnamefont {L.}~\bibnamefont {Bibrzycki}}, \bibinfo {author}
  {\bibfnamefont {A.}~\bibnamefont {Celentano}}, \bibinfo {author}
  {\bibfnamefont {C.}~\bibnamefont {Fern\'andez-Ram\'\i{}rez}}, \ and\ \bibinfo
  {author} {\bibfnamefont {A.~P.}\ \bibnamefont {Szczepaniak}} (\bibinfo
  {collaboration} {JPAC}),\ }\href {\doibase 10.1103/PhysRevD.102.014003}
  {\bibfield  {journal} {\bibinfo  {journal} {Phys. Rev. D}\ }\textbf {\bibinfo
  {volume} {102}},\ \bibinfo {pages} {014003} (\bibinfo {year} {2020})},\
  \Eprint {http://arxiv.org/abs/2005.01617} {arXiv:2005.01617 [hep-ph]}
  \BibitemShut {NoStop}%
\bibitem [{\citenamefont {Mathieu}\ \emph
  {et~al.}(2018{\natexlab{b}})\citenamefont {Mathieu}, \citenamefont {Nys},
  \citenamefont {Fern\'andez-Ram\'\i{}rez}, \citenamefont {Jackura},
  \citenamefont {Pilloni}, \citenamefont {Sherrill}, \citenamefont
  {Szczepaniak},\ and\ \citenamefont {Fox}}]{Mathieu:2018xyc}%
  \BibitemOpen
  \bibfield  {author} {\bibinfo {author} {\bibfnamefont {V.}~\bibnamefont
  {Mathieu}}, \bibinfo {author} {\bibfnamefont {J.}~\bibnamefont {Nys}},
  \bibinfo {author} {\bibfnamefont {C.}~\bibnamefont
  {Fern\'andez-Ram\'\i{}rez}}, \bibinfo {author} {\bibfnamefont
  {A.}~\bibnamefont {Jackura}}, \bibinfo {author} {\bibfnamefont
  {A.}~\bibnamefont {Pilloni}}, \bibinfo {author} {\bibfnamefont
  {N.}~\bibnamefont {Sherrill}}, \bibinfo {author} {\bibfnamefont {A.~P.}\
  \bibnamefont {Szczepaniak}}, \ and\ \bibinfo {author} {\bibfnamefont
  {G.}~\bibnamefont {Fox}} (\bibinfo {collaboration} {JPAC}),\ }\href {\doibase
  10.1103/PhysRevD.97.094003} {\bibfield  {journal} {\bibinfo  {journal} {Phys.
  Rev. D}\ }\textbf {\bibinfo {volume} {97}},\ \bibinfo {pages} {094003}
  (\bibinfo {year} {2018}{\natexlab{b}})},\ \Eprint
  {http://arxiv.org/abs/1802.09403} {arXiv:1802.09403 [hep-ph]} \BibitemShut
  {NoStop}%
\bibitem [{\citenamefont {Albaladejo}\ \emph {et~al.}(2022)\citenamefont
  {Albaladejo} \emph {et~al.}}]{JPAC:2021rxu}%
  \BibitemOpen
  \bibfield  {author} {\bibinfo {author} {\bibfnamefont {M.}~\bibnamefont
  {Albaladejo}} \emph {et~al.} (\bibinfo {collaboration} {JPAC}),\ }\href
  {\doibase 10.1016/j.ppnp.2022.103981} {\bibfield  {journal} {\bibinfo
  {journal} {Prog. Part. Nucl. Phys.}\ }\textbf {\bibinfo {volume} {127}},\
  \bibinfo {pages} {103981} (\bibinfo {year} {2022})},\ \Eprint
  {http://arxiv.org/abs/2112.13436} {arXiv:2112.13436 [hep-ph]} \BibitemShut
  {NoStop}%
\bibitem [{\citenamefont {Pauli}(2022)}]{Pauli:2021gde}%
  \BibitemOpen
  \bibfield  {author} {\bibinfo {author} {\bibfnamefont {P.}~\bibnamefont
  {Pauli}} (\bibinfo {collaboration} {GlueX}),\ }\href {\doibase
  10.31349/SuplRevMexFis.3.0308002} {\bibfield  {journal} {\bibinfo  {journal}
  {Rev. Mex. Fis. Suppl.}\ }\textbf {\bibinfo {volume} {3}},\ \bibinfo {pages}
  {0308002} (\bibinfo {year} {2022})},\ \Eprint
  {http://arxiv.org/abs/2112.05633} {arXiv:2112.05633 [hep-ex]} \BibitemShut
  {NoStop}%
\bibitem [{\citenamefont {Battaglieri}\ \emph {et~al.}(2009)\citenamefont
  {Battaglieri} \emph {et~al.}}]{CLAS:2009ngd}%
  \BibitemOpen
  \bibfield  {author} {\bibinfo {author} {\bibfnamefont {M.}~\bibnamefont
  {Battaglieri}} \emph {et~al.} (\bibinfo {collaboration} {CLAS}),\ }\href
  {\doibase 10.1103/PhysRevD.80.072005} {\bibfield  {journal} {\bibinfo
  {journal} {Phys. Rev. D}\ }\textbf {\bibinfo {volume} {80}},\ \bibinfo
  {pages} {072005} (\bibinfo {year} {2009})},\ \Eprint
  {http://arxiv.org/abs/0907.1021} {arXiv:0907.1021 [hep-ex]} \BibitemShut
  {NoStop}%
\bibitem [{\citenamefont {Strauch}\ \emph {et~al.}(2005)\citenamefont {Strauch}
  \emph {et~al.}}]{CLAS:2005oqk}%
  \BibitemOpen
  \bibfield  {author} {\bibinfo {author} {\bibfnamefont {S.}~\bibnamefont
  {Strauch}} \emph {et~al.} (\bibinfo {collaboration} {CLAS}),\ }\href
  {\doibase 10.1103/PhysRevLett.95.162003} {\bibfield  {journal} {\bibinfo
  {journal} {Phys. Rev. Lett.}\ }\textbf {\bibinfo {volume} {95}},\ \bibinfo
  {pages} {162003} (\bibinfo {year} {2005})},\ \Eprint
  {http://arxiv.org/abs/hep-ex/0508002} {arXiv:hep-ex/0508002} \BibitemShut
  {NoStop}%
\bibitem [{\citenamefont {Battaglieri}\ \emph {et~al.}(2001)\citenamefont
  {Battaglieri} \emph {et~al.}}]{CLAS:2001zxv}%
  \BibitemOpen
  \bibfield  {author} {\bibinfo {author} {\bibfnamefont {M.}~\bibnamefont
  {Battaglieri}} \emph {et~al.} (\bibinfo {collaboration} {CLAS}),\ }\href
  {\doibase 10.1103/PhysRevLett.87.172002} {\bibfield  {journal} {\bibinfo
  {journal} {Phys. Rev. Lett.}\ }\textbf {\bibinfo {volume} {87}},\ \bibinfo
  {pages} {172002} (\bibinfo {year} {2001})},\ \Eprint
  {http://arxiv.org/abs/hep-ex/0107028} {arXiv:hep-ex/0107028} \BibitemShut
  {NoStop}%
\bibitem [{\citenamefont {Aaij}\ \emph {et~al.}(2021)\citenamefont {Aaij} \emph
  {et~al.}}]{LHCb:2021uow}%
  \BibitemOpen
  \bibfield  {author} {\bibinfo {author} {\bibfnamefont {R.}~\bibnamefont
  {Aaij}} \emph {et~al.} (\bibinfo {collaboration} {LHCb}),\ }\href {\doibase
  10.1103/PhysRevLett.127.082001} {\bibfield  {journal} {\bibinfo  {journal}
  {Phys. Rev. Lett.}\ }\textbf {\bibinfo {volume} {127}},\ \bibinfo {pages}
  {082001} (\bibinfo {year} {2021})},\ \Eprint
  {http://arxiv.org/abs/2103.01803} {arXiv:2103.01803 [hep-ex]} \BibitemShut
  {NoStop}%
\bibitem [{LHC(2022)}]{LHCb:2022jad}%
  \BibitemOpen
  \href@noop {} {\  (\bibinfo {year} {2022})},\ \Eprint
  {http://arxiv.org/abs/2210.10346} {arXiv:2210.10346 [hep-ex]} \BibitemShut
  {NoStop}%
\bibitem [{\citenamefont {Mestayer}\ \emph {et~al.}(2000)\citenamefont
  {Mestayer} \emph {et~al.}}]{Mestayer:2000we}%
  \BibitemOpen
  \bibfield  {author} {\bibinfo {author} {\bibfnamefont {M.~D.}\ \bibnamefont
  {Mestayer}} \emph {et~al.},\ }\href {\doibase 10.1016/S0168-9002(00)00151-0}
  {\bibfield  {journal} {\bibinfo  {journal} {Nucl. Instrum. Meth. A}\ }\textbf
  {\bibinfo {volume} {449}},\ \bibinfo {pages} {81} (\bibinfo {year}
  {2000})}\BibitemShut {NoStop}%
\bibitem [{\citenamefont {Smith}\ \emph {et~al.}(1999)\citenamefont {Smith}
  \emph {et~al.}}]{Smith:1999ii}%
  \BibitemOpen
  \bibfield  {author} {\bibinfo {author} {\bibfnamefont {E.~S.}\ \bibnamefont
  {Smith}} \emph {et~al.},\ }\href {\doibase 10.1016/S0168-9002(99)00484-2}
  {\bibfield  {journal} {\bibinfo  {journal} {Nucl. Instrum. Meth. A}\ }\textbf
  {\bibinfo {volume} {432}},\ \bibinfo {pages} {265} (\bibinfo {year}
  {1999})}\BibitemShut {NoStop}%
\bibitem [{\citenamefont {Sober}\ \emph {et~al.}(2000)\citenamefont {Sober}
  \emph {et~al.}}]{Sober:2000we}%
  \BibitemOpen
  \bibfield  {author} {\bibinfo {author} {\bibfnamefont {D.~I.}\ \bibnamefont
  {Sober}} \emph {et~al.},\ }\href {\doibase 10.1016/S0168-9002(99)00784-6}
  {\bibfield  {journal} {\bibinfo  {journal} {Nucl. Instrum. Meth. A}\ }\textbf
  {\bibinfo {volume} {440}},\ \bibinfo {pages} {263} (\bibinfo {year}
  {2000})}\BibitemShut {NoStop}%
\bibitem [{\citenamefont {Whalley}(1989)}]{Whalley:1989mt}%
  \BibitemOpen
  \bibfield  {author} {\bibinfo {author} {\bibfnamefont {M.~R.}\ \bibnamefont
  {Whalley}},\ }\href {\doibase 10.1016/0010-4655(89)90282-8} {\bibfield
  {journal} {\bibinfo  {journal} {Comput. Phys. Commun.}\ }\textbf {\bibinfo
  {volume} {57}},\ \bibinfo {pages} {536} (\bibinfo {year} {1989})}\BibitemShut
  {NoStop}%
\bibitem [{\citenamefont {Schilling}\ \emph {et~al.}(1970)\citenamefont
  {Schilling}, \citenamefont {Seyboth},\ and\ \citenamefont
  {Wolf}}]{Schilling:1969um}%
  \BibitemOpen
  \bibfield  {author} {\bibinfo {author} {\bibfnamefont {K.}~\bibnamefont
  {Schilling}}, \bibinfo {author} {\bibfnamefont {P.}~\bibnamefont {Seyboth}},
  \ and\ \bibinfo {author} {\bibfnamefont {G.~E.}\ \bibnamefont {Wolf}},\
  }\href {\doibase 10.1016/0550-3213(70)90070-2} {\bibfield  {journal}
  {\bibinfo  {journal} {Nucl. Phys. B}\ }\textbf {\bibinfo {volume} {15}},\
  \bibinfo {pages} {397} (\bibinfo {year} {1970})},\ \bibinfo {note} {[Erratum:
  Nucl.Phys.B 18, 332 (1970)]}\BibitemShut {NoStop}%
\bibitem [{GSI()}]{GSIM}%
  \BibitemOpen
  \href@noop {} {\enquote {\bibinfo {title} {Gsim user's guide},}\ }\bibinfo
  {howpublished}
  {\url{https://www.jlab.org/Hall-B/document/gsim/userguide.html}}\BibitemShut
  {NoStop}%
\bibitem [{\citenamefont {Goodfellow}\ \emph {et~al.}(2020)\citenamefont
  {Goodfellow}, \citenamefont {Pouget-Abadie}, \citenamefont {Mirza},
  \citenamefont {Xu}, \citenamefont {Warde-Farley}, \citenamefont {Ozair},
  \citenamefont {Courville},\ and\ \citenamefont
  {Bengio}}]{goodfellow2020generative}%
  \BibitemOpen
  \bibfield  {author} {\bibinfo {author} {\bibfnamefont {I.}~\bibnamefont
  {Goodfellow}}, \bibinfo {author} {\bibfnamefont {J.}~\bibnamefont
  {Pouget-Abadie}}, \bibinfo {author} {\bibfnamefont {M.}~\bibnamefont
  {Mirza}}, \bibinfo {author} {\bibfnamefont {B.}~\bibnamefont {Xu}}, \bibinfo
  {author} {\bibfnamefont {D.}~\bibnamefont {Warde-Farley}}, \bibinfo {author}
  {\bibfnamefont {S.}~\bibnamefont {Ozair}}, \bibinfo {author} {\bibfnamefont
  {A.}~\bibnamefont {Courville}}, \ and\ \bibinfo {author} {\bibfnamefont
  {Y.}~\bibnamefont {Bengio}},\ }in\ \href {\doibase 10.1145/3422622} {\emph
  {\bibinfo {booktitle} {Communications of the ACM}}},\ Vol.~\bibinfo {volume}
  {63}\ (\bibinfo {year} {2020})\ pp.\ \bibinfo {pages} {139--144}\BibitemShut
  {NoStop}%
\bibitem [{\citenamefont {Karras}\ \emph {et~al.}(2019)\citenamefont {Karras},
  \citenamefont {Laine},\ and\ \citenamefont {Aila}}]{karras2019style}%
  \BibitemOpen
  \bibfield  {author} {\bibinfo {author} {\bibfnamefont {T.}~\bibnamefont
  {Karras}}, \bibinfo {author} {\bibfnamefont {S.}~\bibnamefont {Laine}}, \
  and\ \bibinfo {author} {\bibfnamefont {T.}~\bibnamefont {Aila}},\ }in\ \href
  {\doibase 10.1109/TPAMI.2020.2970919} {\emph {\bibinfo {booktitle}
  {Proceedings of the IEEE/CVF conference on computer vision and pattern
  recognition}}},\ Vol.~\bibinfo {volume} {43}\ (\bibinfo {year} {2019})\ pp.\
  \bibinfo {pages} {4401--4410},\ \Eprint {http://arxiv.org/abs/1812.04948}
  {arXiv:1812.04948 [cs.NE]} \BibitemShut {NoStop}%
\bibitem [{\citenamefont {Li}\ \emph {et~al.}(2018)\citenamefont {Li},
  \citenamefont {Pan}, \citenamefont {Wang}, \citenamefont {Yang},\ and\
  \citenamefont {Cambria}}]{li2018generative}%
  \BibitemOpen
  \bibfield  {author} {\bibinfo {author} {\bibfnamefont {Y.}~\bibnamefont
  {Li}}, \bibinfo {author} {\bibfnamefont {Q.}~\bibnamefont {Pan}}, \bibinfo
  {author} {\bibfnamefont {S.}~\bibnamefont {Wang}}, \bibinfo {author}
  {\bibfnamefont {T.}~\bibnamefont {Yang}}, \ and\ \bibinfo {author}
  {\bibfnamefont {E.}~\bibnamefont {Cambria}},\ }\href {\doibase
  10.1016/j.ins.2018.03.050} {\bibfield  {journal} {\bibinfo  {journal} {Inf.
  Sci.}\ }\textbf {\bibinfo {volume} {450}},\ \bibinfo {pages} {301} (\bibinfo
  {year} {2018})}\BibitemShut {NoStop}%
\bibitem [{\citenamefont {Brunner}\ \emph {et~al.}(2018)\citenamefont
  {Brunner}, \citenamefont {Wang}, \citenamefont {Wattenhofer},\ and\
  \citenamefont {Zhao}}]{brunner2018symbolic}%
  \BibitemOpen
  \bibfield  {author} {\bibinfo {author} {\bibfnamefont {G.}~\bibnamefont
  {Brunner}}, \bibinfo {author} {\bibfnamefont {Y.}~\bibnamefont {Wang}},
  \bibinfo {author} {\bibfnamefont {R.}~\bibnamefont {Wattenhofer}}, \ and\
  \bibinfo {author} {\bibfnamefont {S.}~\bibnamefont {Zhao}},\ }in\ \href@noop
  {} {\emph {\bibinfo {booktitle} {2018 IEEE 30th international conference on
  tools with artificial intelligence (ICTAI)}}}\ (\bibinfo {organization}
  {IEEE},\ \bibinfo {year} {2018})\ pp.\ \bibinfo {pages} {786--793},\ \Eprint
  {http://arxiv.org/abs/1809.07575} {arXiv:1809.07575 [cs.SD]} \BibitemShut
  {NoStop}%
\bibitem [{\citenamefont {Clark}\ \emph {et~al.}(2019)\citenamefont {Clark},
  \citenamefont {Donahue},\ and\ \citenamefont
  {Simonyan}}]{clark2019adversarial}%
  \BibitemOpen
  \bibfield  {author} {\bibinfo {author} {\bibfnamefont {A.}~\bibnamefont
  {Clark}}, \bibinfo {author} {\bibfnamefont {J.}~\bibnamefont {Donahue}}, \
  and\ \bibinfo {author} {\bibfnamefont {K.}~\bibnamefont {Simonyan}},\
  }\href@noop {} {\enquote {\bibinfo {title} {Adversarial video generation on
  complex datasets},}\ } (\bibinfo {year} {2019}),\ \Eprint
  {http://arxiv.org/abs/1907.06571} {arXiv:1907.06571 [cs.CV]} \BibitemShut
  {NoStop}%
\bibitem [{\citenamefont {Salimans}\ \emph {et~al.}(2016)\citenamefont
  {Salimans}, \citenamefont {Goodfellow}, \citenamefont {Zaremba},
  \citenamefont {Cheung}, \citenamefont {Radford},\ and\ \citenamefont
  {Chen}}]{salimans2016improved}%
  \BibitemOpen
  \bibfield  {author} {\bibinfo {author} {\bibfnamefont {T.}~\bibnamefont
  {Salimans}}, \bibinfo {author} {\bibfnamefont {I.}~\bibnamefont
  {Goodfellow}}, \bibinfo {author} {\bibfnamefont {W.}~\bibnamefont {Zaremba}},
  \bibinfo {author} {\bibfnamefont {V.}~\bibnamefont {Cheung}}, \bibinfo
  {author} {\bibfnamefont {A.}~\bibnamefont {Radford}}, \ and\ \bibinfo
  {author} {\bibfnamefont {X.}~\bibnamefont {Chen}},\ }in\ \href@noop {} {\emph
  {\bibinfo {booktitle} {Advances in Neural Information Processing Systems 29
  (NIPS 2016)}}},\ Vol.~\bibinfo {volume} {29}\ (\bibinfo {year} {2016})\
  \Eprint {http://arxiv.org/abs/1606.03498} {arXiv:1606.03498 [cs.LG]}
  \BibitemShut {NoStop}%
\bibitem [{\citenamefont {Arora}\ and\ \citenamefont
  {Zhang}(2017)}]{arora2017gans}%
  \BibitemOpen
  \bibfield  {author} {\bibinfo {author} {\bibfnamefont {S.}~\bibnamefont
  {Arora}}\ and\ \bibinfo {author} {\bibfnamefont {Y.}~\bibnamefont {Zhang}},\
  }\href@noop {} {\enquote {\bibinfo {title} {Do gans actually learn the
  distribution? an empirical study},}\ } (\bibinfo {year} {2017}),\ \Eprint
  {http://arxiv.org/abs/1706.08224} {arXiv:1706.08224 [cs.LG]} \BibitemShut
  {NoStop}%
\bibitem [{\citenamefont {Arjovsky}\ and\ \citenamefont
  {Bottou}(2017)}]{arjovsky2017towards}%
  \BibitemOpen
  \bibfield  {author} {\bibinfo {author} {\bibfnamefont {M.}~\bibnamefont
  {Arjovsky}}\ and\ \bibinfo {author} {\bibfnamefont {L.}~\bibnamefont
  {Bottou}},\ }\href@noop {} {\enquote {\bibinfo {title} {Towards principled
  methods for training generative adversarial networks},}\ } (\bibinfo {year}
  {2017}),\ \Eprint {http://arxiv.org/abs/1701.04862} {arXiv:1701.04862
  [stat.ML]} \BibitemShut {NoStop}%
\bibitem [{\citenamefont {Bang}\ and\ \citenamefont
  {Shim}(2021)}]{bang2021mggan}%
  \BibitemOpen
  \bibfield  {author} {\bibinfo {author} {\bibfnamefont {D.}~\bibnamefont
  {Bang}}\ and\ \bibinfo {author} {\bibfnamefont {H.}~\bibnamefont {Shim}},\
  }in\ \href@noop {} {\emph {\bibinfo {booktitle} {Proceedings of the IEEE/CVF
  international conference on computer vision}}}\ (\bibinfo {year} {2021})\
  pp.\ \bibinfo {pages} {2347--2356},\ \Eprint {http://arxiv.org/abs/804.04391}
  {arXiv:804.04391 [cs.CV]} \BibitemShut {NoStop}%
\bibitem [{\citenamefont {Gao}\ \emph {et~al.}(2020)\citenamefont {Gao},
  \citenamefont {H\"oche}, \citenamefont {Isaacson}, \citenamefont {Krause},\
  and\ \citenamefont {Schulz}}]{gao2020event}%
  \BibitemOpen
  \bibfield  {author} {\bibinfo {author} {\bibfnamefont {C.}~\bibnamefont
  {Gao}}, \bibinfo {author} {\bibfnamefont {S.}~\bibnamefont {H\"oche}},
  \bibinfo {author} {\bibfnamefont {J.}~\bibnamefont {Isaacson}}, \bibinfo
  {author} {\bibfnamefont {C.}~\bibnamefont {Krause}}, \ and\ \bibinfo {author}
  {\bibfnamefont {H.}~\bibnamefont {Schulz}},\ }\href {\doibase
  10.1103/PhysRevD.101.076002} {\bibfield  {journal} {\bibinfo  {journal}
  {Phys. Rev. D}\ }\textbf {\bibinfo {volume} {101}},\ \bibinfo {pages}
  {076002} (\bibinfo {year} {2020})},\ \Eprint
  {http://arxiv.org/abs/2001.10028} {arXiv:2001.10028 [hep-ph]} \BibitemShut
  {NoStop}%
\bibitem [{\citenamefont {Otten}\ \emph {et~al.}(2021)\citenamefont {Otten},
  \citenamefont {Caron}, \citenamefont {de~Swart}, \citenamefont {van
  Beekveld}, \citenamefont {Hendriks}, \citenamefont {van Leeuwen},
  \citenamefont {Podareanu}, \citenamefont {Ruiz~de Austri},\ and\
  \citenamefont {Verheyen}}]{otten2021event}%
  \BibitemOpen
  \bibfield  {author} {\bibinfo {author} {\bibfnamefont {S.}~\bibnamefont
  {Otten}}, \bibinfo {author} {\bibfnamefont {S.}~\bibnamefont {Caron}},
  \bibinfo {author} {\bibfnamefont {W.}~\bibnamefont {de~Swart}}, \bibinfo
  {author} {\bibfnamefont {M.}~\bibnamefont {van Beekveld}}, \bibinfo {author}
  {\bibfnamefont {L.}~\bibnamefont {Hendriks}}, \bibinfo {author}
  {\bibfnamefont {C.}~\bibnamefont {van Leeuwen}}, \bibinfo {author}
  {\bibfnamefont {D.}~\bibnamefont {Podareanu}}, \bibinfo {author}
  {\bibfnamefont {R.}~\bibnamefont {Ruiz~de Austri}}, \ and\ \bibinfo {author}
  {\bibfnamefont {R.}~\bibnamefont {Verheyen}},\ }\href {\doibase
  10.1038/s41467-021-22616-z} {\bibfield  {journal} {\bibinfo  {journal}
  {Nature Commun.}\ }\textbf {\bibinfo {volume} {12}},\ \bibinfo {pages} {2985}
  (\bibinfo {year} {2021})},\ \Eprint {http://arxiv.org/abs/1901.00875}
  {arXiv:1901.00875 [hep-ph]} \BibitemShut {NoStop}%
\bibitem [{\citenamefont {Paganini}\ \emph
  {et~al.}(2018{\natexlab{a}})\citenamefont {Paganini}, \citenamefont
  {de~Oliveira},\ and\ \citenamefont {Nachman}}]{paganini2018calogan}%
  \BibitemOpen
  \bibfield  {author} {\bibinfo {author} {\bibfnamefont {M.}~\bibnamefont
  {Paganini}}, \bibinfo {author} {\bibfnamefont {L.}~\bibnamefont
  {de~Oliveira}}, \ and\ \bibinfo {author} {\bibfnamefont {B.}~\bibnamefont
  {Nachman}},\ }\href {\doibase 10.1103/PhysRevD.97.014021} {\bibfield
  {journal} {\bibinfo  {journal} {Phys. Rev. D}\ }\textbf {\bibinfo {volume}
  {97}},\ \bibinfo {pages} {014021} (\bibinfo {year} {2018}{\natexlab{a}})},\
  \Eprint {http://arxiv.org/abs/1712.10321} {arXiv:1712.10321 [hep-ex]}
  \BibitemShut {NoStop}%
\bibitem [{\citenamefont {Alanazi}\ \emph
  {et~al.}(2021{\natexlab{a}})\citenamefont {Alanazi}, \citenamefont {Sato},
  \citenamefont {Liu}, \citenamefont {Melnitchouk}, \citenamefont
  {Ambrozewicz}, \citenamefont {Hauenstein}, \citenamefont {Kuchera},
  \citenamefont {Pritchard}, \citenamefont {Robertson}, \citenamefont
  {Strauss}, \citenamefont {Velasco},\ and\ \citenamefont
  {Li}}]{ijcai2021p293}%
  \BibitemOpen
  \bibfield  {author} {\bibinfo {author} {\bibfnamefont {Y.}~\bibnamefont
  {Alanazi}}, \bibinfo {author} {\bibfnamefont {N.}~\bibnamefont {Sato}},
  \bibinfo {author} {\bibfnamefont {T.}~\bibnamefont {Liu}}, \bibinfo {author}
  {\bibfnamefont {W.}~\bibnamefont {Melnitchouk}}, \bibinfo {author}
  {\bibfnamefont {P.}~\bibnamefont {Ambrozewicz}}, \bibinfo {author}
  {\bibfnamefont {F.}~\bibnamefont {Hauenstein}}, \bibinfo {author}
  {\bibfnamefont {M.~P.}\ \bibnamefont {Kuchera}}, \bibinfo {author}
  {\bibfnamefont {E.}~\bibnamefont {Pritchard}}, \bibinfo {author}
  {\bibfnamefont {M.}~\bibnamefont {Robertson}}, \bibinfo {author}
  {\bibfnamefont {R.}~\bibnamefont {Strauss}}, \bibinfo {author} {\bibfnamefont
  {L.}~\bibnamefont {Velasco}}, \ and\ \bibinfo {author} {\bibfnamefont
  {Y.}~\bibnamefont {Li}},\ }in\ \href {\doibase 10.24963/ijcai.2021/293}
  {\emph {\bibinfo {booktitle} {Proceedings of the Thirtieth International
  Joint Conference on Artificial Intelligence, {IJCAI-21}}}}\ (\bibinfo {year}
  {2021})\ p.\ \bibinfo {pages} {2126}\BibitemShut {NoStop}%
\bibitem [{\citenamefont {Hashemi}\ \emph {et~al.}()\citenamefont {Hashemi},
  \citenamefont {Amin}, \citenamefont {Datta}, \citenamefont {Olivito},\ and\
  \citenamefont {Pierini}}]{hashemi2019lhc}%
  \BibitemOpen
  \bibfield  {author} {\bibinfo {author} {\bibfnamefont {B.}~\bibnamefont
  {Hashemi}}, \bibinfo {author} {\bibfnamefont {N.}~\bibnamefont {Amin}},
  \bibinfo {author} {\bibfnamefont {K.}~\bibnamefont {Datta}}, \bibinfo
  {author} {\bibfnamefont {D.}~\bibnamefont {Olivito}}, \ and\ \bibinfo
  {author} {\bibfnamefont {M.}~\bibnamefont {Pierini}},\ }\href@noop {}
  {\enquote {\bibinfo {title} {{LHC analysis-specific datasets with generative
  adversarial networks}},}\ }\Eprint {http://arxiv.org/abs/1901.05282}
  {arXiv:1901.05282 [hep-ex]} \BibitemShut {NoStop}%
\bibitem [{\citenamefont {Butter}\ \emph {et~al.}(2019)\citenamefont {Butter},
  \citenamefont {Plehn},\ and\ \citenamefont {Winterhalder}}]{butter2019GAN}%
  \BibitemOpen
  \bibfield  {author} {\bibinfo {author} {\bibfnamefont {A.}~\bibnamefont
  {Butter}}, \bibinfo {author} {\bibfnamefont {T.}~\bibnamefont {Plehn}}, \
  and\ \bibinfo {author} {\bibfnamefont {R.}~\bibnamefont {Winterhalder}},\
  }\href {\doibase 10.21468/SciPostPhys.7.6.075} {\bibfield  {journal}
  {\bibinfo  {journal} {SciPost Phys.}\ }\textbf {\bibinfo {volume} {7}},\
  \bibinfo {pages} {075} (\bibinfo {year} {2019})},\ \Eprint
  {http://arxiv.org/abs/1907.03764} {arXiv:1907.03764 [hep-ph]} \BibitemShut
  {NoStop}%
\bibitem [{\citenamefont {Di~Sipio}\ \emph {et~al.}(2020)\citenamefont
  {Di~Sipio}, \citenamefont {Faucci}, \citenamefont {Ketabchi},\ and\
  \citenamefont {Palazzo}}]{DiSipio:2019imz}%
  \BibitemOpen
  \bibfield  {author} {\bibinfo {author} {\bibfnamefont {R.}~\bibnamefont
  {Di~Sipio}}, \bibinfo {author} {\bibfnamefont {G.~M.}\ \bibnamefont
  {Faucci}}, \bibinfo {author} {\bibfnamefont {H.~S.}\ \bibnamefont
  {Ketabchi}}, \ and\ \bibinfo {author} {\bibfnamefont {S.}~\bibnamefont
  {Palazzo}},\ }\href {\doibase 10.1007/JHEP08(2019)110} {\bibfield  {journal}
  {\bibinfo  {journal} {JHEP}\ }\textbf {\bibinfo {volume} {08}},\ \bibinfo
  {pages} {110} (\bibinfo {year} {2020})},\ \Eprint
  {http://arxiv.org/abs/1903.02433} {arXiv:1903.02433 [hep-ex]} \BibitemShut
  {NoStop}%
\bibitem [{\citenamefont {Paganini}\ \emph
  {et~al.}(2018{\natexlab{b}})\citenamefont {Paganini}, \citenamefont
  {de~Oliveira},\ and\ \citenamefont {Nachman}}]{paganini2018accelerating}%
  \BibitemOpen
  \bibfield  {author} {\bibinfo {author} {\bibfnamefont {M.}~\bibnamefont
  {Paganini}}, \bibinfo {author} {\bibfnamefont {L.}~\bibnamefont
  {de~Oliveira}}, \ and\ \bibinfo {author} {\bibfnamefont {B.}~\bibnamefont
  {Nachman}},\ }\href {\doibase 10.1103/PhysRevLett.120.042003} {\bibfield
  {journal} {\bibinfo  {journal} {Phys. Rev. Lett.}\ }\textbf {\bibinfo
  {volume} {120}},\ \bibinfo {pages} {042003} (\bibinfo {year}
  {2018}{\natexlab{b}})},\ \Eprint {http://arxiv.org/abs/1705.02355}
  {arXiv:1705.02355 [hep-ex]} \BibitemShut {NoStop}%
\bibitem [{\citenamefont {de~Oliveira}\ \emph {et~al.}(2017)\citenamefont
  {de~Oliveira}, \citenamefont {Paganini},\ and\ \citenamefont
  {Nachman}}]{de2017learning}%
  \BibitemOpen
  \bibfield  {author} {\bibinfo {author} {\bibfnamefont {L.}~\bibnamefont
  {de~Oliveira}}, \bibinfo {author} {\bibfnamefont {M.}~\bibnamefont
  {Paganini}}, \ and\ \bibinfo {author} {\bibfnamefont {B.}~\bibnamefont
  {Nachman}},\ }\href {\doibase 10.1007/s41781-017-0004-6} {\bibfield
  {journal} {\bibinfo  {journal} {Comput. Softw. Big Sci.}\ }\textbf {\bibinfo
  {volume} {1}},\ \bibinfo {pages} {4} (\bibinfo {year} {2017})},\ \Eprint
  {http://arxiv.org/abs/1701.05927} {arXiv:1701.05927 [stat.ML]} \BibitemShut
  {NoStop}%
\bibitem [{\citenamefont {Musella}\ and\ \citenamefont
  {Pandolfi}(2018)}]{musella2018fast}%
  \BibitemOpen
  \bibfield  {author} {\bibinfo {author} {\bibfnamefont {P.}~\bibnamefont
  {Musella}}\ and\ \bibinfo {author} {\bibfnamefont {F.}~\bibnamefont
  {Pandolfi}},\ }\href {\doibase 10.1007/s41781-018-0015-y} {\bibfield
  {journal} {\bibinfo  {journal} {Comput. Softw. Big Sci.}\ }\textbf {\bibinfo
  {volume} {2}},\ \bibinfo {pages} {8} (\bibinfo {year} {2018})},\ \Eprint
  {http://arxiv.org/abs/1805.00850} {arXiv:1805.00850 [hep-ex]} \BibitemShut
  {NoStop}%
\bibitem [{\citenamefont {Bellagente}\ \emph {et~al.}(2020)\citenamefont
  {Bellagente}, \citenamefont {Butter}, \citenamefont {Kasieczka},
  \citenamefont {Plehn},\ and\ \citenamefont
  {Winterhalder}}]{bellagente2020gan}%
  \BibitemOpen
  \bibfield  {author} {\bibinfo {author} {\bibfnamefont {M.}~\bibnamefont
  {Bellagente}}, \bibinfo {author} {\bibfnamefont {A.}~\bibnamefont {Butter}},
  \bibinfo {author} {\bibfnamefont {G.}~\bibnamefont {Kasieczka}}, \bibinfo
  {author} {\bibfnamefont {T.}~\bibnamefont {Plehn}}, \ and\ \bibinfo {author}
  {\bibfnamefont {R.}~\bibnamefont {Winterhalder}},\ }\href {\doibase
  10.21468/SciPostPhys.8.4.070} {\bibfield  {journal} {\bibinfo  {journal}
  {SciPost Phys.}\ }\textbf {\bibinfo {volume} {8}},\ \bibinfo {pages} {070}
  (\bibinfo {year} {2020})},\ \Eprint {http://arxiv.org/abs/1912.00477}
  {arXiv:1912.00477 [hep-ph]} \BibitemShut {NoStop}%
\bibitem [{\citenamefont {Alanazi}\ \emph
  {et~al.}(2021{\natexlab{b}})\citenamefont {Alanazi}, \citenamefont {Sato},
  \citenamefont {Ambrozewicz}, \citenamefont {Hiller-Blin}, \citenamefont
  {Melnitchouk}, \citenamefont {Battaglieri}, \citenamefont {Liu},\ and\
  \citenamefont {Li}}]{ijcai2021p588}%
  \BibitemOpen
  \bibfield  {author} {\bibinfo {author} {\bibfnamefont {Y.}~\bibnamefont
  {Alanazi}}, \bibinfo {author} {\bibfnamefont {N.}~\bibnamefont {Sato}},
  \bibinfo {author} {\bibfnamefont {P.}~\bibnamefont {Ambrozewicz}}, \bibinfo
  {author} {\bibfnamefont {A.}~\bibnamefont {Hiller-Blin}}, \bibinfo {author}
  {\bibfnamefont {W.}~\bibnamefont {Melnitchouk}}, \bibinfo {author}
  {\bibfnamefont {M.}~\bibnamefont {Battaglieri}}, \bibinfo {author}
  {\bibfnamefont {T.}~\bibnamefont {Liu}}, \ and\ \bibinfo {author}
  {\bibfnamefont {Y.}~\bibnamefont {Li}},\ }in\ \href {\doibase
  10.24963/ijcai.2021/588} {\emph {\bibinfo {booktitle} {Proceedings of the
  Thirtieth International Joint Conference on Artificial Intelligence,
  {IJCAI-21}}}}\ (\bibinfo {year} {2021})\ pp.\ \bibinfo {pages}
  {4286--4293}\BibitemShut {NoStop}%
\bibitem [{\citenamefont {Mao}\ \emph {et~al.}(2017)\citenamefont {Mao},
  \citenamefont {Li}, \citenamefont {Xie}, \citenamefont {Lau}, \citenamefont
  {Wang},\ and\ \citenamefont {Smolley}}]{mao2017squares}%
  \BibitemOpen
  \bibfield  {author} {\bibinfo {author} {\bibfnamefont {X.}~\bibnamefont
  {Mao}}, \bibinfo {author} {\bibfnamefont {Q.}~\bibnamefont {Li}}, \bibinfo
  {author} {\bibfnamefont {H.}~\bibnamefont {Xie}}, \bibinfo {author}
  {\bibfnamefont {R.~Y.~K.}\ \bibnamefont {Lau}}, \bibinfo {author}
  {\bibfnamefont {Z.}~\bibnamefont {Wang}}, \ and\ \bibinfo {author}
  {\bibfnamefont {S.~P.}\ \bibnamefont {Smolley}},\ }in\ \href {\doibase
  10.1109/ICCV.2017.304} {\emph {\bibinfo {booktitle} {Proceedings of the 2017
  IEEE International Conference on Computer Vision (ICCV)}}}\ (\bibinfo {year}
  {2017})\ pp.\ \bibinfo {pages} {2813--2821},\ \Eprint
  {http://arxiv.org/abs/1611.04076} {arXiv:1611.04076 [cs.CV]} \BibitemShut
  {NoStop}%
\bibitem [{\citenamefont {Mirza}\ and\ \citenamefont
  {Osindero}(2014)}]{mirza2014conditional}%
  \BibitemOpen
  \bibfield  {author} {\bibinfo {author} {\bibfnamefont {M.}~\bibnamefont
  {Mirza}}\ and\ \bibinfo {author} {\bibfnamefont {S.}~\bibnamefont
  {Osindero}},\ }\href@noop {} {\  (\bibinfo {year} {2014})},\ \Eprint
  {http://arxiv.org/abs/1411.1784} {arXiv:1411.1784 [cs.LG]} \BibitemShut
  {NoStop}%
\bibitem [{\citenamefont {Kingma}\ and\ \citenamefont
  {Ba}(2014)}]{kingma2014adam}%
  \BibitemOpen
  \bibfield  {author} {\bibinfo {author} {\bibfnamefont {D.~P.}\ \bibnamefont
  {Kingma}}\ and\ \bibinfo {author} {\bibfnamefont {J.}~\bibnamefont {Ba}},\
  }in\ \href@noop {} {\emph {\bibinfo {booktitle} {3rd International Conference
  for Learning Representations (ICLR)}}}\ (\bibinfo {year} {2014})\ \Eprint
  {http://arxiv.org/abs/1412.6980} {arXiv:1412.6980 [cs.LG]} \BibitemShut
  {NoStop}%
\end{thebibliography}%

\end{document}